\documentclass[twocolumn,showpacs, nofootinbib]{revtex4-1}

\usepackage[percent]{overpic}
\usepackage{amssymb}
\usepackage{ctable}
\usepackage{enumitem}
\usepackage{epsfig}
\usepackage{epstopdf}
\usepackage{graphicx}
\usepackage{multirow}
\usepackage{subfigure}
\usepackage{tikz}
\usepackage{amsmath}
\usepackage{placeins}

\graphicspath{{figures/}}

\begin{document}

\title{A Topological Framework for Local Structure Analysis in Condensed Matter}

\author{Emanuel A. Lazar$^1$, Jian Han$^1$, David J. Srolovitz$^{1,2}$}
\affiliation
{$^1$Department of Materials Science and Engineering\\
 $^2$Department of Mechanical Engineering and Applied Mechanics\\University of Pennsylvania, Philadelphia, Pennsylvania 19104, USA.}
\date{\today}

\begin{abstract}
Physical systems are frequently modeled as sets of points in space, each representing the position of an atom, molecule, or mesoscale particle.  As many properties of such systems depend on the underlying ordering of their constituent particles, understanding that structure is a primary objective of condensed matter research.  Although perfect crystals are fully described by a set of translation and basis vectors, real-world materials are never perfect, as thermal vibrations and defects introduce significant deviation from ideal order.  Meanwhile, liquids and glasses present yet more complexity.  A complete understanding of structure thus remains a central, open problem.  Here we propose a unified mathematical framework, based on the topology of the Voronoi cell of a particle, for classifying local structure in ordered and disordered systems that is powerful and practical.  We explain the underlying reason why this topological description of local structure is better suited for structural analysis than continuous descriptions.  We demonstrate the connection of this approach to the behavior of physical systems and explore how crystalline structure is compromised at elevated temperatures.  We also illustrate potential applications to identifying defects in plastically deformed polycrystals at high temperatures, automating analysis of complex structures, and characterizing general disordered systems.  
\end{abstract}

\maketitle


Condensed matter systems are often abstracted as large sets of points in space, each representing the position of an atom, molecule, or mesoscale particle.  Two challenges frequently encountered when studying systems at this scale are classifying and identifying local structure.  Simulation studies of nucleation, crystallization, and melting, for example, as well as those of defect migration and transformation, require a precise understanding of which particles are associated with which phases, and which are associated with defects.  As these systems are abstracted as large point sets, these dual challenges of classifying and identifying structure reduce to ones of understanding arrangements of points in space.  

A primary difficulty in classifying structure in spatial point sets arises from a tension between a desire for completeness and the necessity for practicality.  The local neighborhood of a particle within an ensemble of particles can be completely described by a list of relative positions of each of its neighbors.  However, while such a list of coordinates is complete in some sense, this raw data provides little direct insight, leaving us wanting for a practical and more illuminating description.  This tension is often mediated by the choice of an ``order parameter'', which distills structural data into a single number or vector, and which is constructed to be both informative and computationally tractable \cite{truskett2000towards}. 

A central limitation of conventional order parameters is exhibited in degeneracies that arise in describing neighborhoods that are structurally distinct but which map to identical order-parameter values.  Some order parameters classify particles in face-centered cubic (FCC) and body-centered cubic (BCC) crystals identically, while others classify particles in FCC and hexagonal close-packed (HCP) crystals identically \cite{stukowski2012structure}.  Similarly, particles located near defects in a low-temperature crystal can have order-parameter values identical to those of particles in a high-temperature defect-free crystal.  These degeneracies point to an inherent incompleteness in such order-parameter classifications of local structure.  Consequently, different order parameters are necessary to study different systems \cite{truskett2000towards, stukowski2012structure}. 
\begin{figure}[h]
\includegraphics[width=1.0\columnwidth]{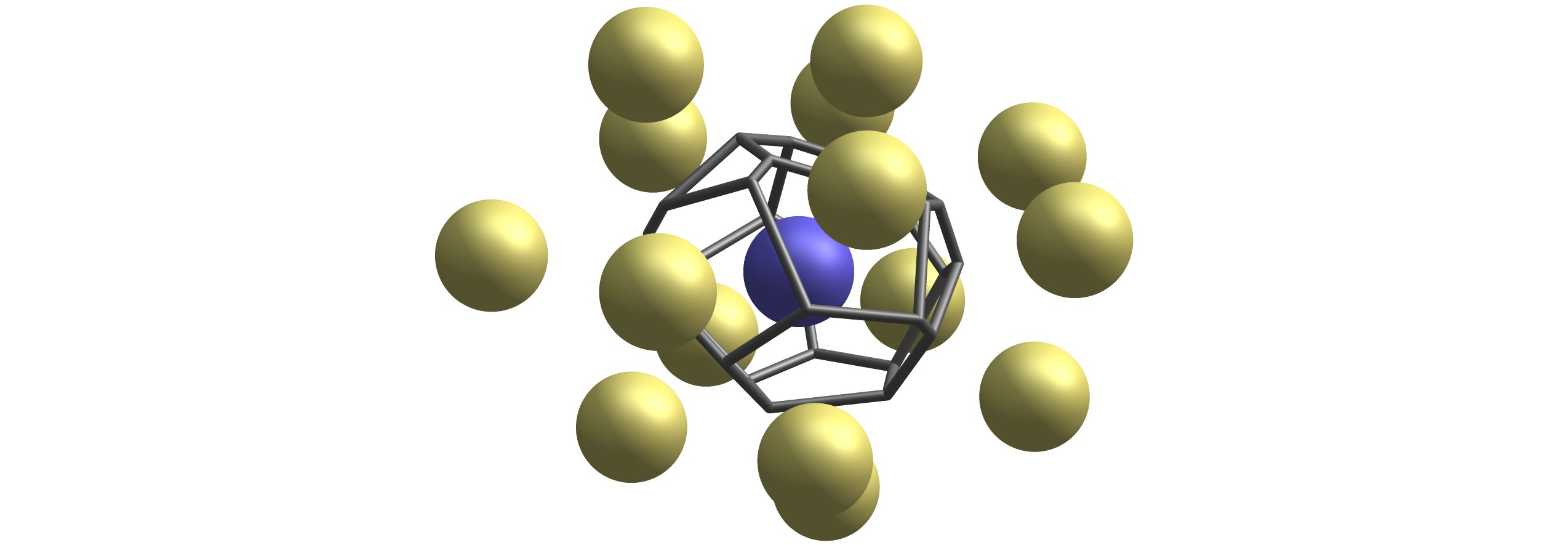} 
\caption{The frame of a Voronoi cell of a central particle (blue), surrounded by its nearest neighbors (gold).  The topology of the Voronoi cell captures structural information about the local neighborhood.\label{example_cell}}
\end{figure}

In this paper we propose a mathematical framework to classify local structure that avoids much of the degeneracy encountered in other approaches and which, therefore, is equally applicable to all ordered and disordered systems of particles.  More specifically, the local structure around a particle is classified using the topology of its Voronoi cell (see Fig.~\ref{example_cell}).  Families of Voronoi cell topologies are constructed by considering those topologies that can result from infinitesimal perturbations of an ideal structure.  We demonstrate that this classification scheme is consistent with the manner in which local ordering changes in high-temperature single crystals as the temperature is raised toward their melting points.  We highlight a potential use of this approach for visualizing defects in crystalline solids at high temperatures, and contrast it with previous methods.  We then demonstrate an application of this approach to the automated analysis of the evolution of complex structures, where conventional methods are often inadequate.  Finally, we show an application in which this approach is used to provide robust statistical-structural descriptors for characterizing disordered systems.  

\section{Theory}
\subsection{The Configuration Space of Local Structure}

\begin{figure}
\includegraphics[width=0.96\columnwidth]{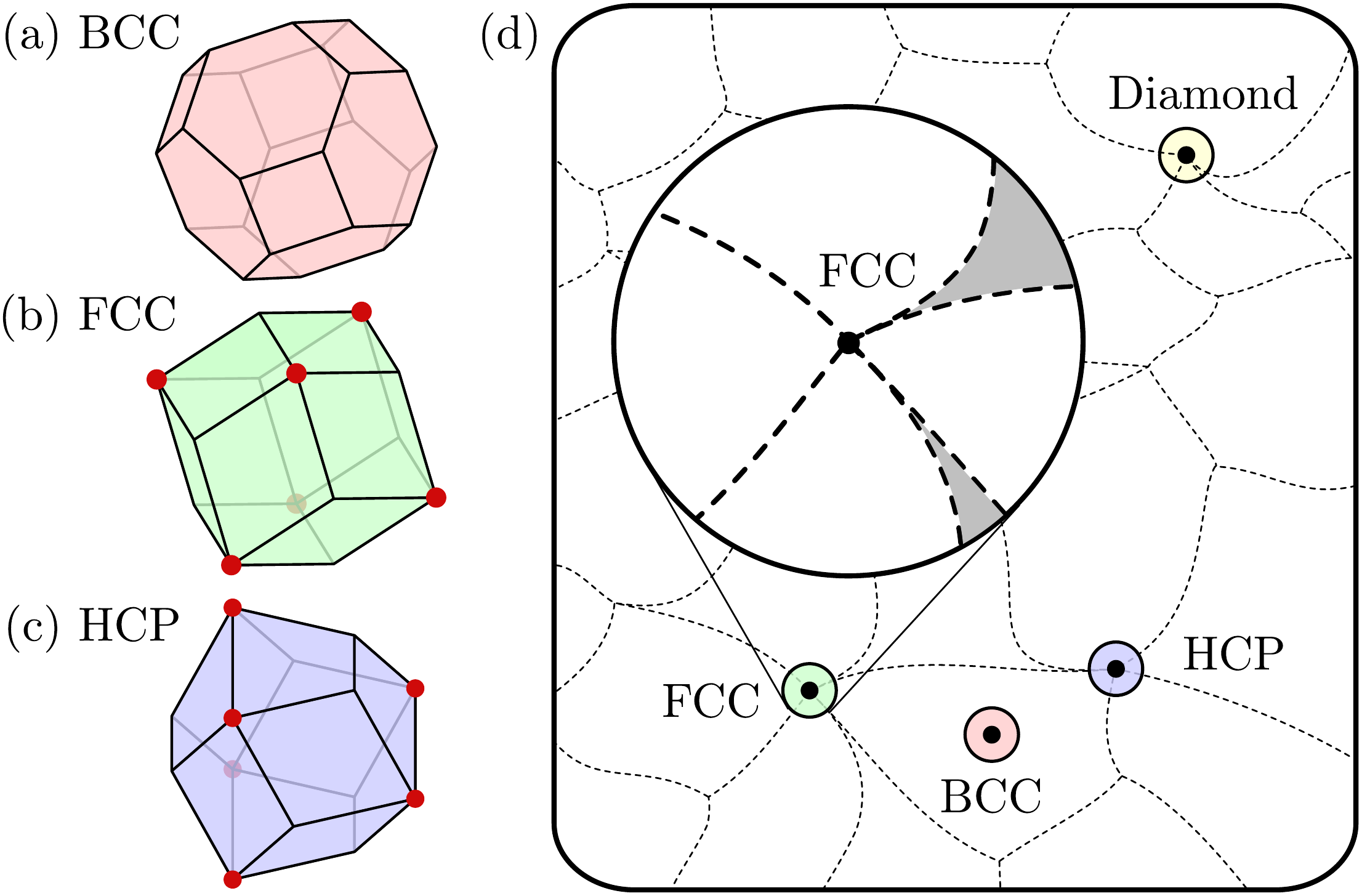} 
\caption{(a-c) Voronoi cells of particles in BCC, FCC, and HCP crystals.  Vertices at which more than four Voronoi cells meet are marked by red circles.  Small perturbations of the particle positions result in topological changes near these vertices. (d) Schematic of $\mathcal C(n)$, the space of all possible configurations of $n$ neighbors.  This space can be divided into regions on which the Voronoi cell topology is constant.  The topology of a point that lies on the boundaries of multiple regions is unstable, and infinitesimal perturbations of the neighbors will result in a change of topology.  The inset shows the neighborhood around $\mathbf x_{\text{FCC}}$; unshaded regions indicate primary types, while shaded ones indicate secondary types.\label{mapping}}
\end{figure}

A deeper understanding of local structure can be developed through consideration of all possible arrangements of neighbors of a central particle.  The local neighborhood of a particle within an ensemble of particles can be completely described by a vector of relative positions of its $n$ nearest neighbors: $\mathbf x = (\mathbf r_1, \mathbf r_2, ..., \mathbf r_n)$, where $\mathbf r_i$ is the relative position of the $i$th neighbor of a central particle.  For suitably large $n$, any question about the local neighborhood of a particle can be answered through complete knowledge of $\mathbf x$.  We use $\mathcal C(n)$ to denote the configuration space of all possible arrangements of $n$ nearest neighbors:  
\begin{equation}
\mathcal C(n) = \{(\mathbf r_1, \mathbf r_2, ..., \mathbf r_n) : \mathbf r_i \in \mathbb{R}^3 \}.
\label{cnequation}
\end{equation}
Each point in $\mathcal C(n)$ thus corresponds to a specific local arrangements of particles.  Figure \ref{mapping}(d) provides a schematic of $\mathcal C(n)$ and highlights points corresponding to local arrangements of BCC, FCC, HCP, and diamond structures.  As defined in Eq.~\ref{cnequation}, the dimension of $\mathcal C(n)$ is $3n$; ignoring rotations and scaling reduces the dimension of $\mathcal C(n)$ by 4.  Ignoring permutations of the $n$ neighbors and disallowing multiplicities further changes the geometry and topology of $\mathcal C(n)$, but not its dimension. 

Order parameters can be thought of as functions that map $\mathcal C(n)$ to a lower-dimensional order-parameter space; order-parameter spaces most commonly used are $\mathbb{R}^d$, where $d$ is substantially smaller than $3n$.  Each choice of order parameter results in a different subdivision of $\mathcal C(n)$ into regions on which that order parameter is constant; for real-valued continuous functions, these regions are commonly known as level sets.  To help understand the  degeneracy observed in continuous order-parameter methods, consider that for every continuous mapping $\phi$ from an unbounded high-dimensional space to a lower-dimensional space, there exist points $\mathbf x_1, \mathbf x_2$ arbitrarily far apart, but for which $\phi(\mathbf x_1)$ and $\phi(\mathbf x_2)$ are identical \cite{2015landweber}.  The continuity of an order parameter thus entails the kind of degeneracy highlighted above.  In contrast, discrete order parameters are not subject to this limitation, as distances between points with identical order-parameter values can be bounded.  This motivates the question of how to reasonably subdivide $\mathcal C(n)$.  We now show that Voronoi topology offers one such approach.

\subsection{Voronoi Topology}

For a fixed set of particles, the Voronoi cell of a central particle is the region of real space closer to that particle than to any other \cite{voronoi1908nouvelles}.  Figure \ref{example_cell} illustrates a central particle, its Voronoi cell, and fifteen neighboring particles.  Two particles whose Voronoi cells share a face are called neighbors.  We identify two Voronoi cells as having the same topology, or {\bf topological type}, if there exists a one-to-one correspondence between their sets of faces that preserves adjacency.  

The topology of the Voronoi cell of a particle describes the manner in which neighbors of a particle are arranged relative to it.  An $n$-sided face, for example, indicates a pair of particles which have $n$ neighbors in common.  The topology of a Voronoi cell thus provides a robust description of how neighbors are arranged not only relative to a central particle, but also to one another.  In this sense, it is a good description of local structure.  

Voronoi cell topology also provides a natural decomposition of $\mathcal C(n)$ into regions in which the Voronoi cell topology is constant, as illustrated in Fig.~\ref{mapping}(d).  We consider this decomposition natural because it allows us to coarse-grain the effects of small perturbations on local structure.  Small perturbations of the particle coordinates correspond to small displacements in $\mathcal C(n)$, and since the Voronoi topology does not change for almost all points under small perturbations, these small perturbations, which are often unimportant, are naturally ignored without the introduction of an artificial cut-off \cite{stukowski2012structure}.  

Voronoi cell topology was first introduced by Bernal and others to study the atomic structure of liquids \cite{bernal1959geometrical, rahman1966liquid, finney1970random0}, and has been subsequently applied to study a wide range of condensed matter systems, including random sphere packings \cite{finney1970random0,bernal1967random}, finite-temperature crystals \cite{hsu1979interaction}, and metallic glasses \cite{sheng2006atomic}.  In those studies, however, the topology of a cell was characterized by counting its types of faces (e.g., triangles and quadrilaterals), though it ignored the way in which those faces are arranged.  While this limited description has been used to study some aspects of crystallization \cite{tanemura1977geometrical}, it cannot distinguish particles whose local environments are FCC from those whose local environments are HCP, as both Voronoi cells have twelve four-sided faces.  In previous work \cite{lazar2011evolution, 2012lazar}, the authors have shown how to use a graph-tracing algorithm introduced by Weinberg \cite{1966weinberg1} to efficiently compute strings which encode a complete description of the Voronoi cell topology; see Methods for further details.

A second limitation arising in traditional Voronoi approaches results from abrupt changes in topology due to small geometric perturbations.  Consider, for example, that Voronoi cells of particles in FCC and HCP crystals are topologically unstable -- since some vertices are shared by more than four Voronoi cells (see Fig.~\ref{mapping}), infinitesimal perturbations of the particle positions, such as those arising from non-hydrostatic strain or thermal vibrations, will change their topology \cite{troadec1998statistics}.  This problem has been sufficiently challenging to limit the utility of conventional Voronoi approaches in studying even slightly perturbed crystal structures \cite{stukowski2012structure}.  This problem can be solved through the classification of topological types described in the following section.

\subsection{Theory of $\lambda$-types}

In this section we show how topological types can be classified using the approach developed in the previous two sections.  On a basic level, every arrangement of neighbors relative to a central particle can be described by its Voronoi cell topology.  Families of topological types associated with a particular structure can then be defined as sets of types obtained through infinitesimal perturbations of that structure.  This classification scheme enables a description of the effects of small strains and thermal vibrations on local structure, and provides a robust framework suitable for theoretical and numerical analysis. 

Every local arrangement of neighbors $\lambda$ is described by a distinct point $\mathbf x_{\lambda}$ in $\mathcal C(n)$, and subsequently corresponds to a unique Voronoi cell topology $V[\mathbf x_{\lambda}]$.  For example, if $\lambda = $ BCC, then $\mathbf x_{\lambda} = \mathbf x_{\text{BCC}}$ describes a particle that has the same local environment as a particle in a perfect BCC crystal; its Voronoi cell topology $V[\mathbf x_{\text{BCC}}]$ is the truncated octahedron, illustrated in Fig.~\ref{mapping}(a). 

A suitable distance function on $\mathcal C(n)$ allows us to define sets of topological types associated with infinitesimal perturbations as follows.  We let $B_{\epsilon}(\mathbf x)$ be a ball of radius $\epsilon$ centered at $\mathbf x$.  This region of $\mathcal C(n)$ corresponds to configurations obtained through small perturbations of a particle and its neighbors, where $\epsilon$ controls the magnitude of such a perturbation.  The set of topological types obtained from all possible perturbations of $\mathbf x$ with magnitude smaller than $\epsilon$ is denoted $V[B_{\epsilon}(\mathbf x)]$.  We define the family of topological types associated with infinitesimal perturbations of $\lambda$ as the limiting set:
\begin{equation}  
\mathcal F_{\lambda} := \lim_{\epsilon \to 0} V\left[B_{\epsilon}(\mathbf x_{\lambda})\right].
\label{families}
\end{equation}  
In more physical terms, $\mathcal F_{\lambda}$ is the set of all topological types that can be obtained through arbitrarily small perturbations of a central particle and its neighbors.  The Voronoi cell topology of points in the interior of a region in $\mathcal C(n)$ remains unchanged by small perturbations.  In contrast, points such as $\mathbf x_{\text{FCC}}$ and $\mathbf x_{\text{HCP}}$ are located at the boundaries of multiple regions, and small perturbations entail a change in Voronoi cell topology.  Thus, $\mathcal F_{\text{FCC}}$ and $\mathcal F_{\text{HCP}}$ consist of multiple topological types, whereas $\mathcal F_{\text{BCC}}$, located in the interior of a region, consists of a single type.  If a topological type is in $\mathcal F_{\lambda}$, then we say that it is a $\lambda$-type.  Note that a topological type can belong to multiple families; this indeterminacy will be considered below.  This classification of $\lambda$-types allows us to account for topological instability without modifying the sample data by collapsing edges or faces using {\it ad hoc} criteria (e.g., cut-offs) \cite{stukowski2012structure,hsu1979interaction}. 

Among $\lambda$-types, a further distinction can be drawn based on the manner in which the Voronoi cell topology subdivides $\mathcal C(n)$.  Using a suitable volume measure Vol, we define the ideal frequency $f_{\lambda}(\tau)$ of a topological type $\tau$ relative to $\mathbf x_{\lambda}$ as follows:
\begin{equation}  
f_{\lambda}(\tau) := \lim_{\epsilon \to 0} \frac{\text{Vol}\left(V^{-1}[\tau]\cap B_{\epsilon}(\mathbf x_{\lambda})\right)}{\text{Vol}\left(B_{\epsilon}(\mathbf x_{\lambda})\right)},
\label{primarytypes}
\end{equation}  
where $V^{-1}[\tau]$ is the set of points in $\mathcal C(n)$ whose Voronoi cell have topology $\tau$.  If $f_{\lambda}(\tau)>0$, we call $\tau$ a primary $\lambda$-type; if $f_{\lambda}(\tau)=0$, we call it a secondary $\lambda$-type.  The inset in Fig.~\ref{mapping}(d) highlights a number of regions incident with $\mathbf x_{\text{FCC}}$.  Some of those regions meet $\mathbf x_{\text{FCC}}$ at finite solid angles; therefore, their fractional volumes within an $\epsilon$-ball converge to positive values as $\epsilon \to 0$; these are primary FCC-types.  In contrast, fractional volumes tend to zero as $\epsilon \to 0$ for other regions which meet $\mathbf x_{\text{FCC}}$ at cusps; these are secondary FCC-types. 

\begin{figure}
\includegraphics[width=0.95\columnwidth]{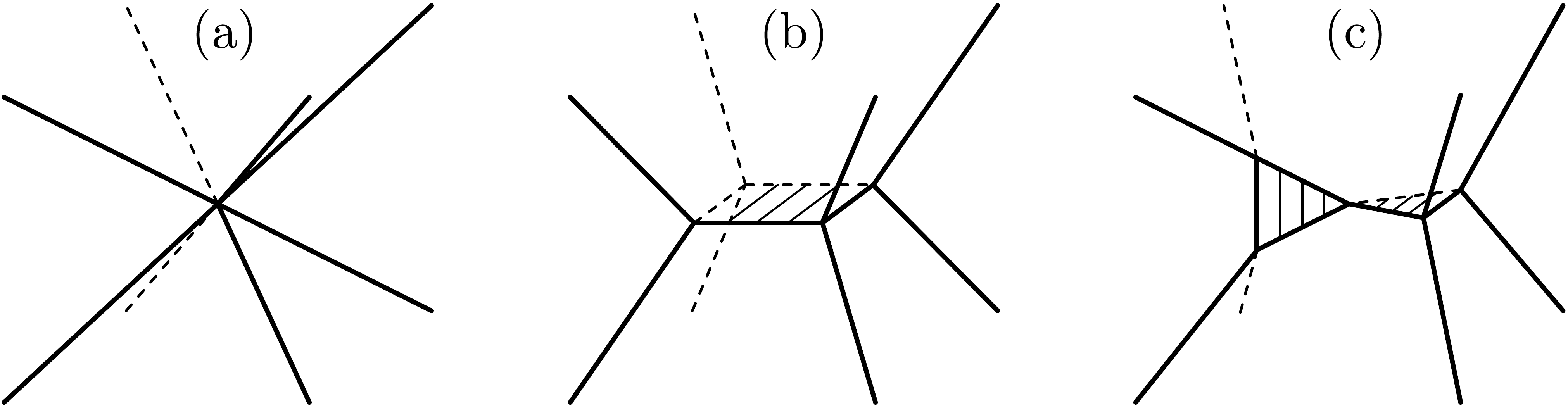} 
\vspace{1mm}
\caption{(a) An unstable vertex shared by six Voronoi cells in FCC or HCP crystals; such vertices are marked by red circles in Figs.~\ref{mapping}(b) and (c).  A small perturbation will cause the vertex to resolve into either (b) a four-sided face, or (c) a pair of adjacent triangular faces; these resolutions are associated with primary and secondary types, respectively.\label{transition}}
\end{figure}

The distinction between primary and secondary types appears to result from the manner in which topological instabilities resolve when perturbed.  To illustrate this distinction, Fig.~\ref{transition}(a) shows an unstable vertex shared by six Voronoi cells in FCC or HCP crystals; such vertices are marked by red circles in Figs.~\ref{mapping}(b) and (c).  Figure \ref{transition}(b) depicts the most common manner in which such an unstable vertex resolves upon small perturbations of neighboring particles \cite{troadec1998statistics}.  In this resolution, a new four-sided face is formed between two non-adjacent Voronoi cells; all unstable vertices resolve in this manner in primary types.  A less common resolution can also occur as a result of cooperative motion of neighboring particles.  In this resolution, depicted in Fig.~\ref{transition}(c), two triangular faces are created \cite{tanemura1977geometrical}; secondary types can include such resolutions. 

Determining $\mathcal F_{\lambda}$ is feasible through consideration of all possible ways in which unstable vertices can resolve.  For example, the ideal FCC Voronoi cell, illustrated in Fig.~\ref{mapping}(b), has six unstable vertices.  In primary types, each such vertex resolves in a manner illustrated in Fig.~\ref{transition}(b), in one of three directions.  More specifically, the unstable vertex can transform in such a way that the central cell gains a square face, or else gains an edge in one of two directions.  We consider all possible combinations of these resolutions over the six unstable vertices, and calculate the topological types of the resulting polyhedra using the algorithm developed in \cite{2012lazar}; a total of 44 distinct topological types occur in this manner.  In secondary types, unstable vertices can also resolve in the manner illustrated in Fig.~\ref{transition}(c), or else remain unstable.  An additional 6250 topological types can occur in this manner.  A similar approach can be followed to determine $\mathcal F_{\lambda}$ for other structures.  Additional details can be found in the supplementary material.


\section{Finite-Temperature Crystals}

\begin{figure*}
\includegraphics[clip=true, height=0.225\linewidth]{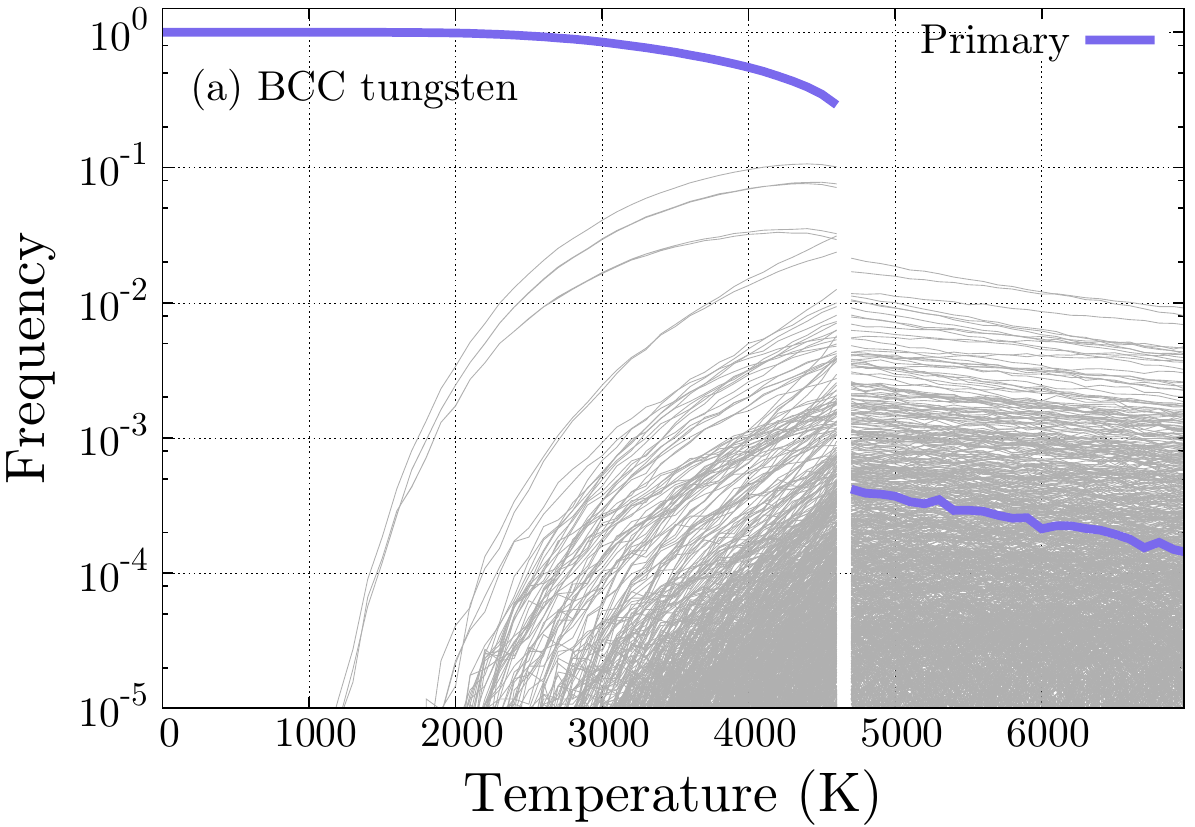} \hfill
\includegraphics[clip=true, height=0.225\linewidth]{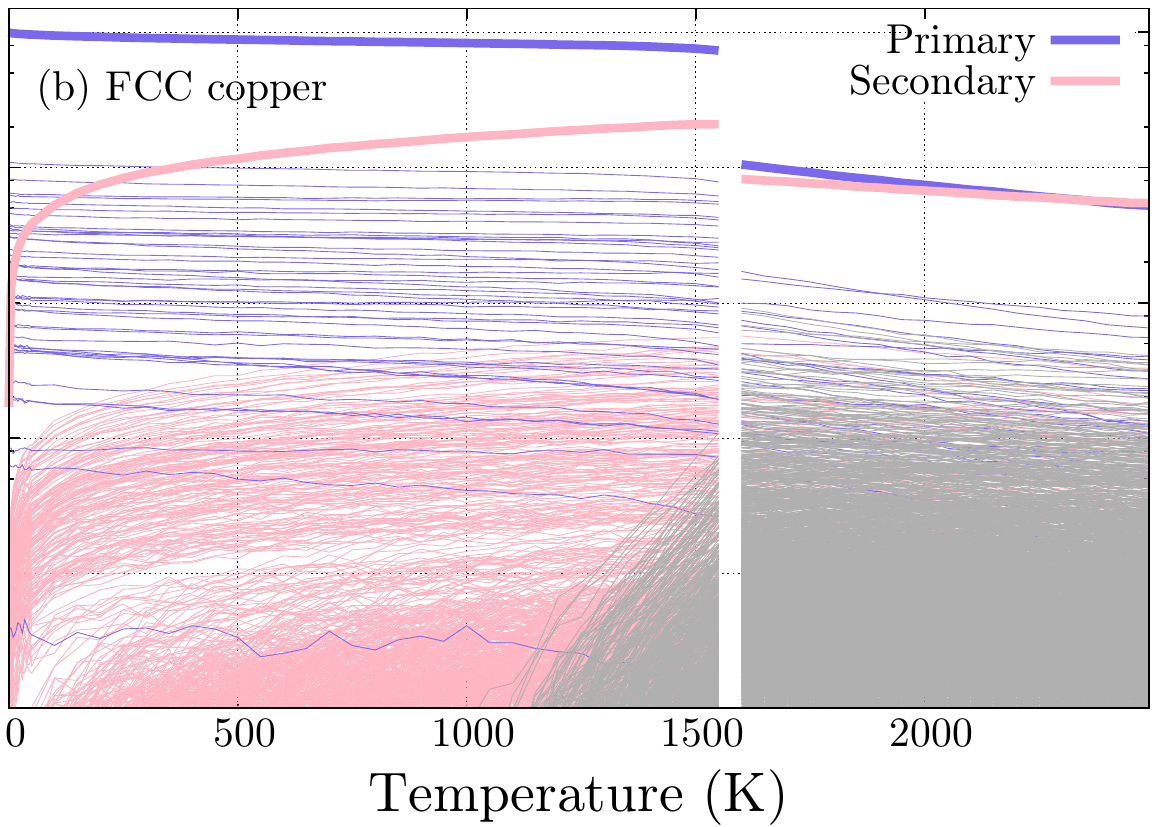} \hfill
\includegraphics[clip=true, height=0.225\linewidth]{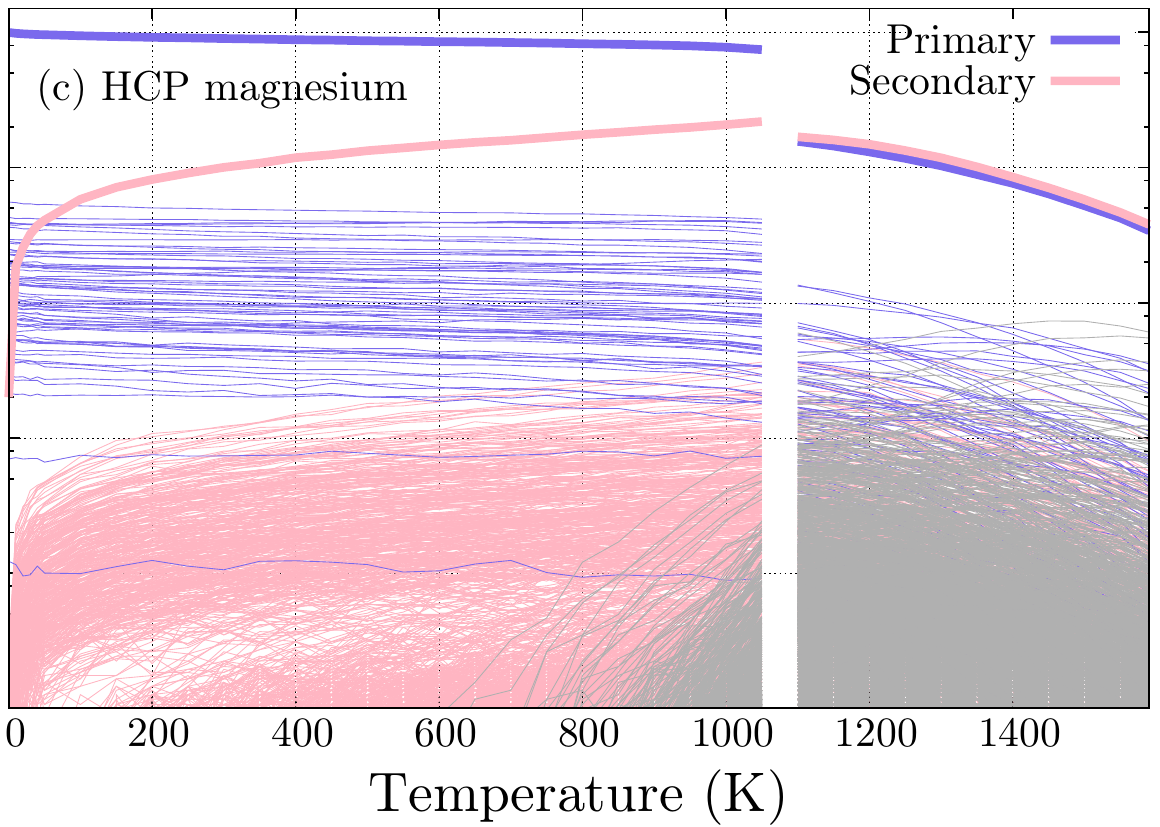}
\caption{Frequencies of all $\lambda$-types and all non-$\lambda$-types that appear in the single-crystal (a) BCC tungsten, (b) FCC copper, and (c) HCP magnesium as a function of temperature upon heating from $T = 0$ to 150\% of the bulk melting temperature.  Blue curves indicate primary $\lambda$-types, pink curves secondary $\lambda$-types, and grey curves non-$\lambda$-types.  Thick curves indicate the sum of all frequencies of the corresponding color; note that there is only one primary $\lambda$-type and no secondary $\lambda$-types for BCC.}
\label{all_types}
\end{figure*}

The proposed distinction between primary and secondary types is supported by atomistic simulation.  We studied the atomic structure of three model materials, BCC tungsten \cite{ackland1987improved}, FCC copper \cite{mishin2001structural}, and HCP magnesium \cite{sun2006crystal}, at elevated temperatures using molecular dynamics (MD) in the $NPT$ ensemble \cite{plimpton1995fast}.  Simulated systems contained 1,024,000, 1,372,000, and 1,029,600 atoms, respectively, in periodic supercells.  In each simulation, a defect-free crystal was heated from $T=0$ in increments of 50 K and equilibrated for 50 ps at each temperature.  Figure \ref{all_types} shows how the distribution of topological types changes with temperature; each curve indicates the frequency of a single topological type.

Types can be grouped according to the shape of their frequency curves.  Frequencies of one group of types approach finite values as $T \to 0$, and change very little with temperature (blue curves in Fig.~\ref{all_types}).  Frequencies of a second group rapidly approach zero as $T \to 0$ (pink curves).  Types of a third group only appear at high temperatures (grey curves).  Remarkably, these groups correspond exactly to the sets of primary $\lambda$, secondary $\lambda$, and non-$\lambda$-types for each system, as enumerated using the analysis of instabilities, described above.  The theory of $\lambda$-types is thus consistent with observed results and suggests a new approach for analyzing thermal effects.

One noteworthy feature of Fig.~\ref{all_types} is the similarity between FCC and HCP, in contrast to BCC.  These general behaviors appear to depend primarily on the crystal structure rather than on bonding particulars. Indeed, preliminary investigations show that when atoms in BCC, FCC, and HCP crystals are perturbed from their equilibrium positions with a Gaussian noise whose magnitude scales with temperature (i.e., an Einstein model \cite{einstein1907planck}), the distribution of topological types changes in a manner similar to that observed in Fig.~\ref{all_types}.  

A second notable feature is the total frequency of $\lambda$-types in the liquid phases of the three systems.  In liquid tungsten, the unique BCC-type accounts for less than $0.05\%$ of all atoms just above $T_{\text{m}}$, where $T_{\text{m}}$ is the bulk melting temperature.  In contrast, liquid copper consists of roughly 20\% FCC-types, and liquid magnesium consists of roughly 30\% HCP-types, just above $T_{\text{m}}$.  These structural data are relevant in studying physical processes such as crystallization \cite{kawasaki2010formation} and melting \cite{bai2008ring}.  

A third feature that stands out is the high fraction of $\lambda$-types in the FCC and HCP samples at temperatures just below melting.  In FCC, 94\% of all atoms are classified as having FCC local structure just below melting; in HCP this number is 96\%.  At 0.85$T_{\text{m}}$, these numbers are over 99\% in both systems.  This suggests a natural use of $\lambda$-types for identifying structure in highly-perturbed atomistic systems, such as those at high temperatures.  

We next consider several applications of the proposed approach to illustrate some of its unique features.

\section{Defect Visualization}

\begin{figure}[h]
\begin{tabular}[c]{c}
\fbox{\includegraphics[width=0.955\columnwidth]{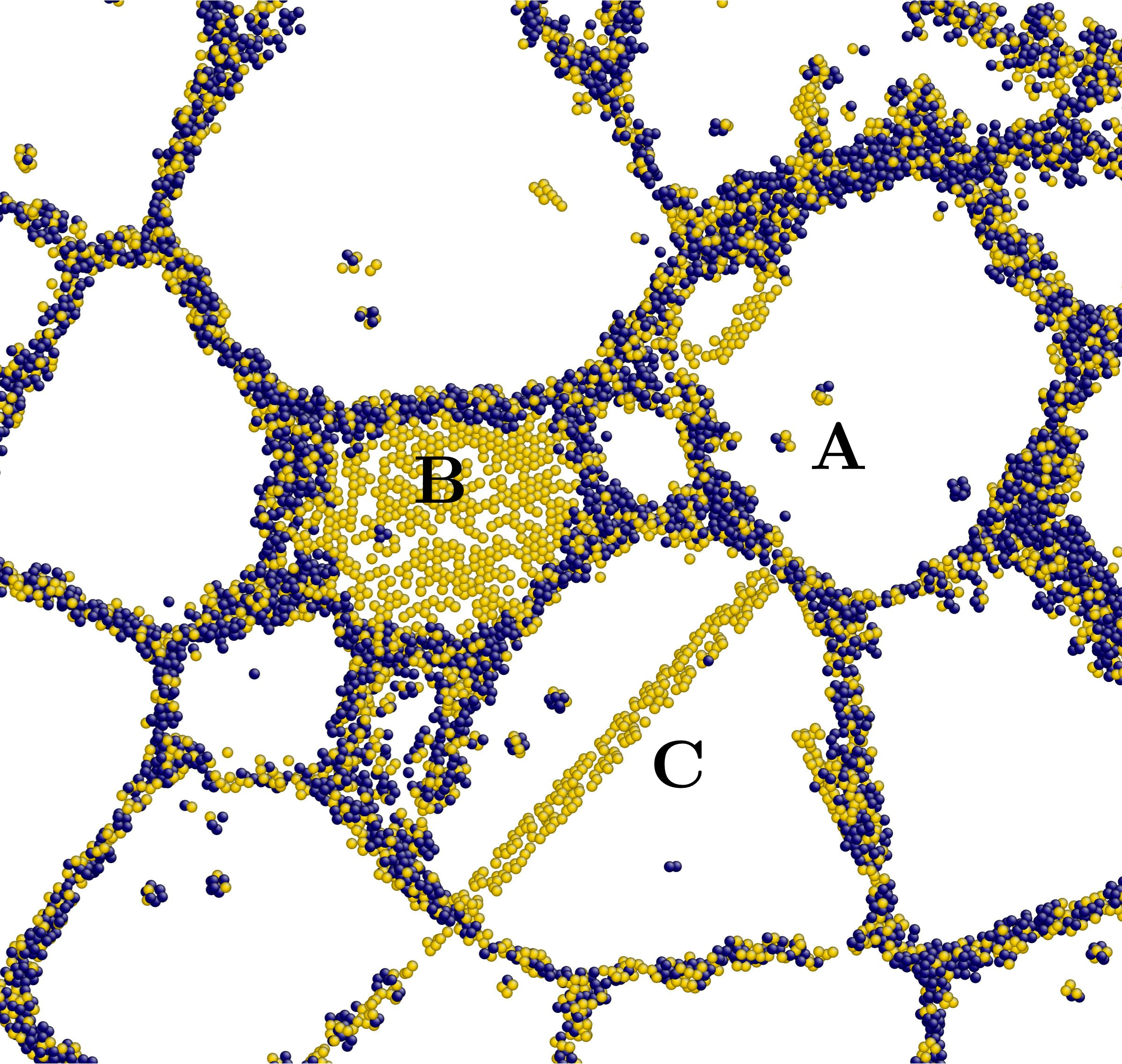}}\vspace{1mm}\\  
(a) Polycrystal cross-section \vspace{1mm}\\ 
\end{tabular}
\begin{tabular}[c]{cc}
\fbox{\includegraphics[width=0.46\columnwidth]{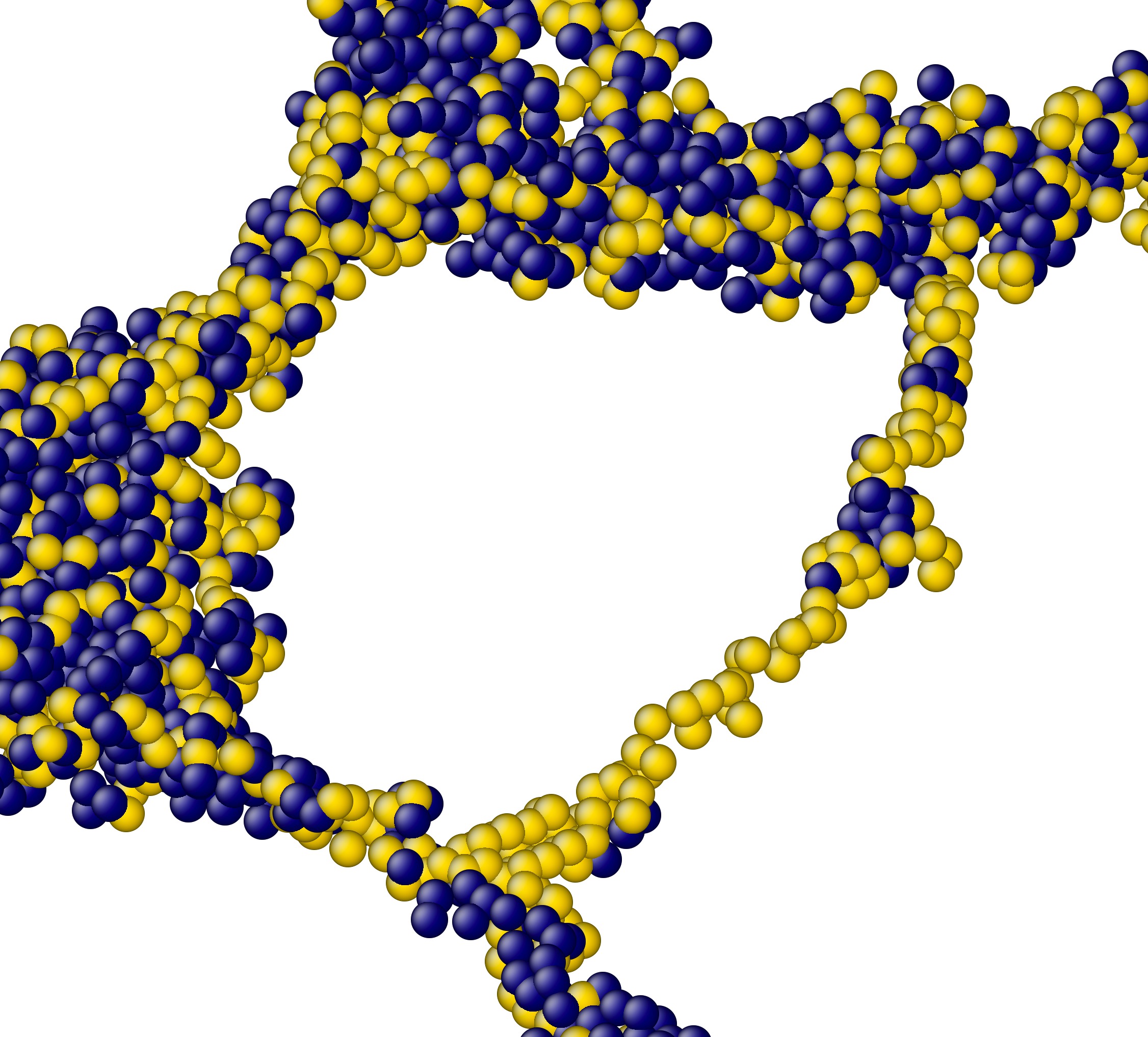}} \quad & 
\fbox{\includegraphics[width=0.46\columnwidth]{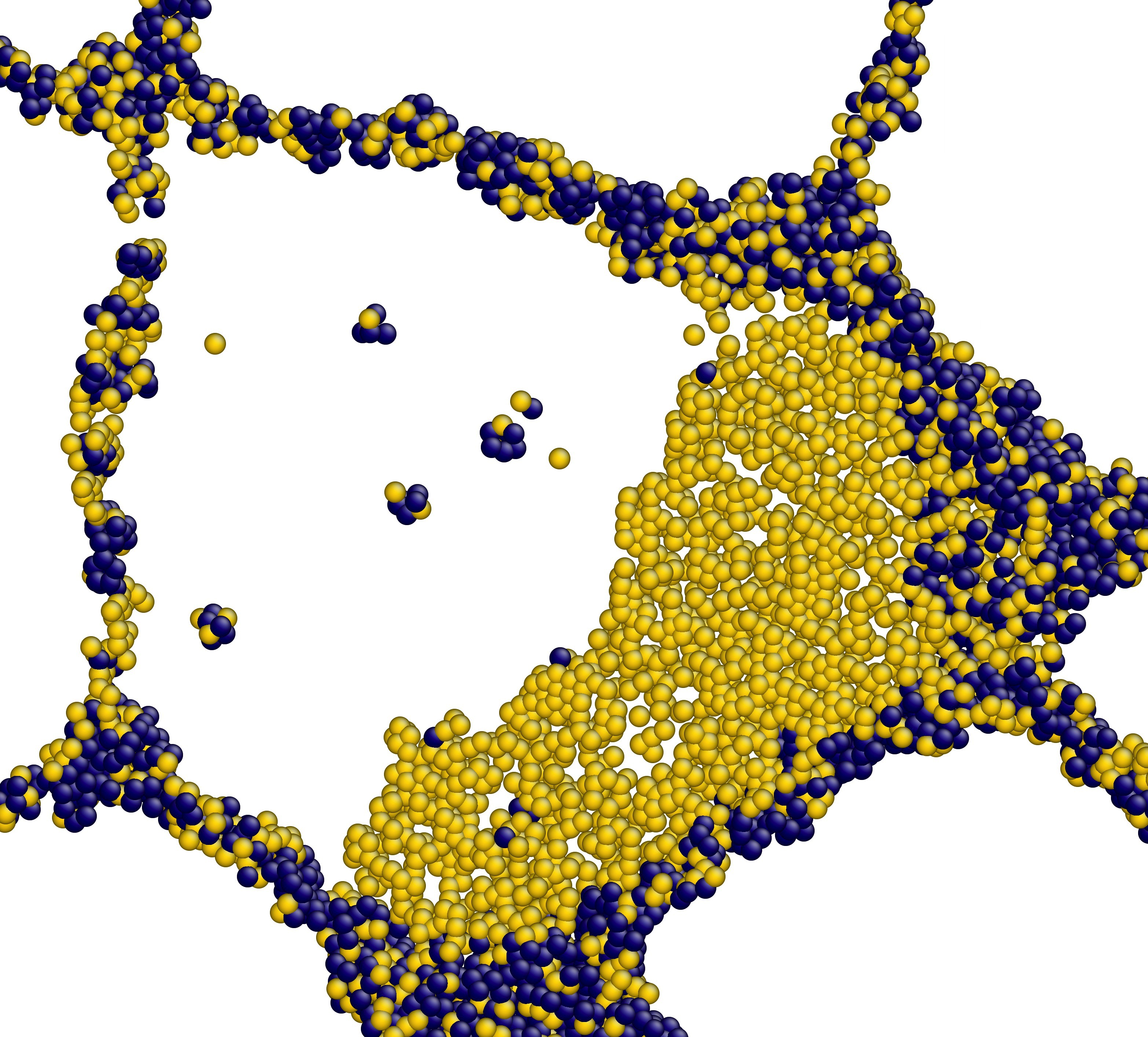}}\vspace{1mm}\\      
(b) Dislocation & (c) Stacking fault\\
\end{tabular}
\caption{Polycrystalline aluminum at 938 K ($0.9 T_{\text{m}}$); the width of each cross-section is 2 nm. Atoms that are FCC-types are not shown for clarity.  Of the ones remaining, those that are HCP-types are shown in gold and all other atoms are shown in dark blue.  Grain boundaries are seen as a network of non-FCC-types (dark blue and gold atoms).  In cross-section (a), defects are labeled as follow: vacancies {\bf A}, twin boundary {\bf B}, and  stacking fault {\bf C}. Cross-sections (b) and (c) show magnified images of a dislocation and stacking fault.\label{polycrystal}}
\end{figure}

As noted, the high frequencies of $\lambda$-types in single crystals, even at extremely high temperatures, suggests their use for visualization of local structure in atomic systems.  Figure \ref{polycrystal} shows thin cross-sections from an FCC aluminum polycrystal prepared using MD \cite{mishin1999interatomic}.  The sample was obtained by annealing a microstructure obtained through simulated grain-growth \cite{lazar2011more}, plastically deforming it at low temperature, and then thermalizing it at $0.9 T_{\text{m}}$.  In these figures, atoms that are FCC-types are not shown for clarity; among the remaining atoms, those that are HCP-types are shown in gold, and all other atoms are shown in dark blue.  
\setlength{\fboxsep}{0pt}

In Fig.~\ref{polycrystal}(a), grain boundaries can be identified as a network of non-FCC-type atoms.  Vacancies ({\bf A}) can be identified within the grain interiors as small clusters of non-FCC-types; as only a thin cross-section of the material is shown, not all atoms adjacent to these defects appear in the figure.  A twin boundary ({\bf B}) and stacking fault ({\bf C}) can be identified as one and two layers of gold-colored (HCP, non-FCC type) atoms, respectively.  Figures \ref{polycrystal}(b) and (c) show magnified images of a dislocation and a stacking fault inclined at a low angle relative to the cross-section plane.  

As noted earlier, individual Voronoi topologies can belong to multiple families; we use the term {\bf indeterminate type} to refer to such cases.  This indeterminacy complicates the visualization procedure suggested here, as many types in $\mathcal F_{\text{HCP}}$ also belong to $\mathcal F_{\text{FCC}}$.  For this reason some atoms that belong to the twin boundaries and stacking faults are not seen in Figs.~\ref{polycrystal}(a) and (c).  This shortcoming can be easily addressed within the topological framework, and is discussed in the supplementary material.

The utility of the proposed topological framework for local structure classification and identification is useful for finite-temperature simulations of atomic systems containing defects. Of particular interest are the many phenomena which are only of interest at high temperature, such as dislocation climb \cite{hirth1982theory}, interface thermal fluctuation \cite{foiles2006computation}, and defect kinetics under irradiation conditions \cite{fu2004multiscale}.  Conventional visualization methods require quenching or temporally averaging a sample prior to crystal defect analysis \cite{stukowski2012structure}. In general, we do not know whether this ``tampering'' with the data leads to significant discrepancies between the observed structures and the actual finite-temperature ones.  Remarkably, the proposed approach identifies all defects in Fig.~\ref{polycrystal} and does not erroneously identify any bulk atoms as being adjacent to defects, all at very high temperature and without quenching or time averaging.  The topological approach thus provides a natural method for identifying and visualizing local structure that involves no {\it ad hoc} cut-off parameters and which is robust at high temperatures.  This opens new opportunity for {\it in situ} structure analysis of atomic simulations at temperature, potentially identifying new high-temperature mechanisms and defect structures.

\section{Comparison with Other Methods}

Although a complete survey of existing methods for analyzing structure in high-temperature atomic systems is beyond the scope of this introductory paper, we briefly consider how visualization using $\lambda$-types compares with three of the most frequently-used order parameters: centrosymmetry \cite{kelchner1998dislocation}, bond-angle analysis \cite{ackland2006applications}, and adaptive common-neighbor analysis \cite{stukowski2012structure, honeycutt1987molecular}.

\begin{figure}[h]
\begin{tabular}[c]{cc}
\includegraphics[width=0.47\columnwidth]{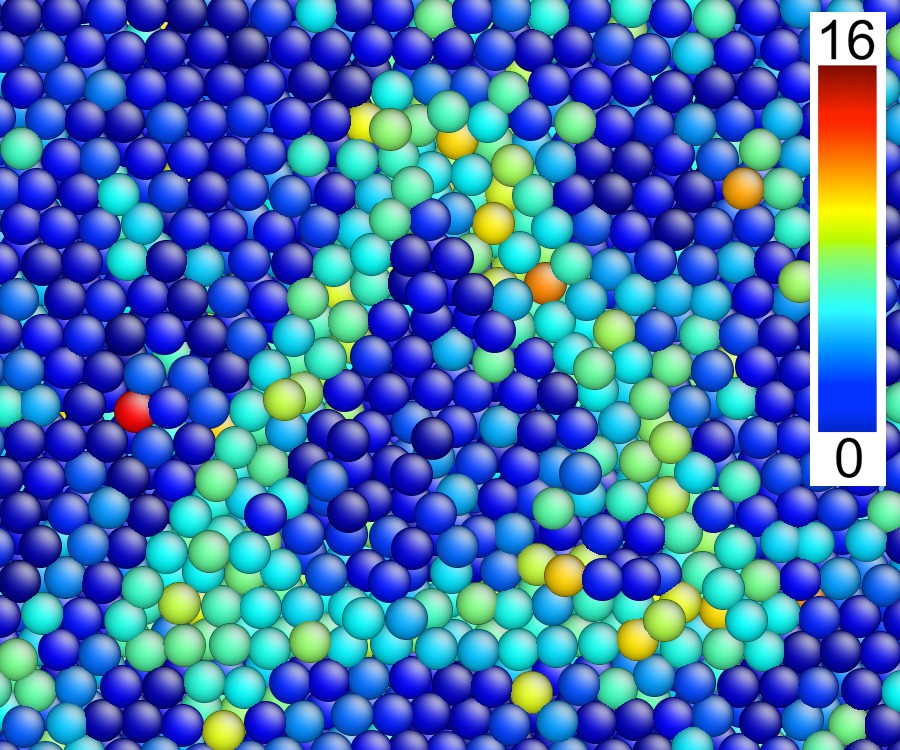} \quad &
\includegraphics[width=0.47\columnwidth]{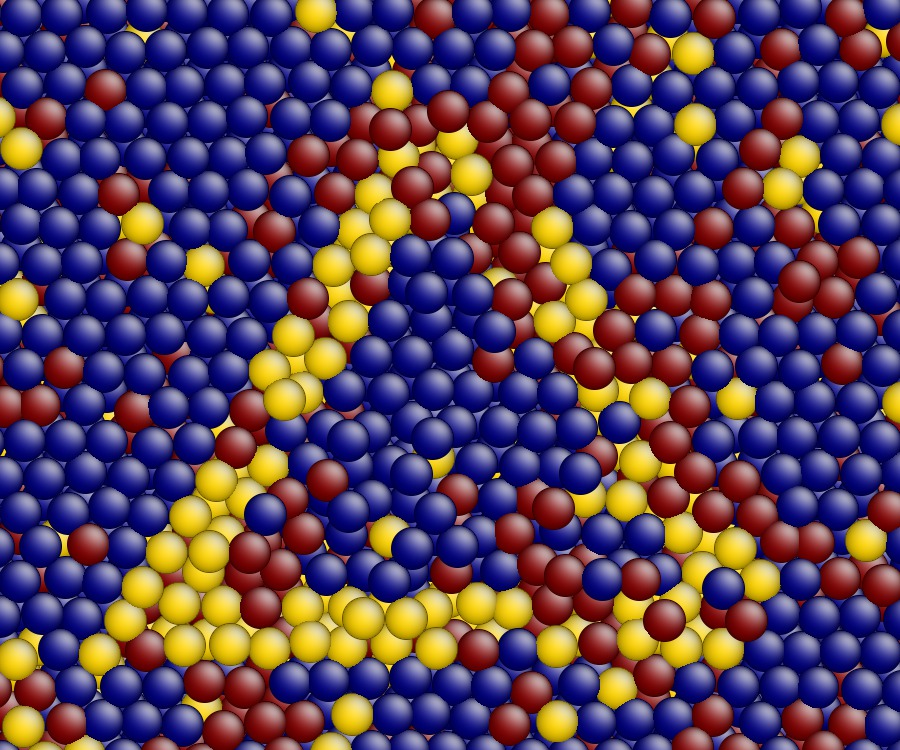} \\
(a) Centrosymmetry  & (b) Bond-angle analysis \vspace{2mm} \\
\includegraphics[width=0.47\columnwidth]{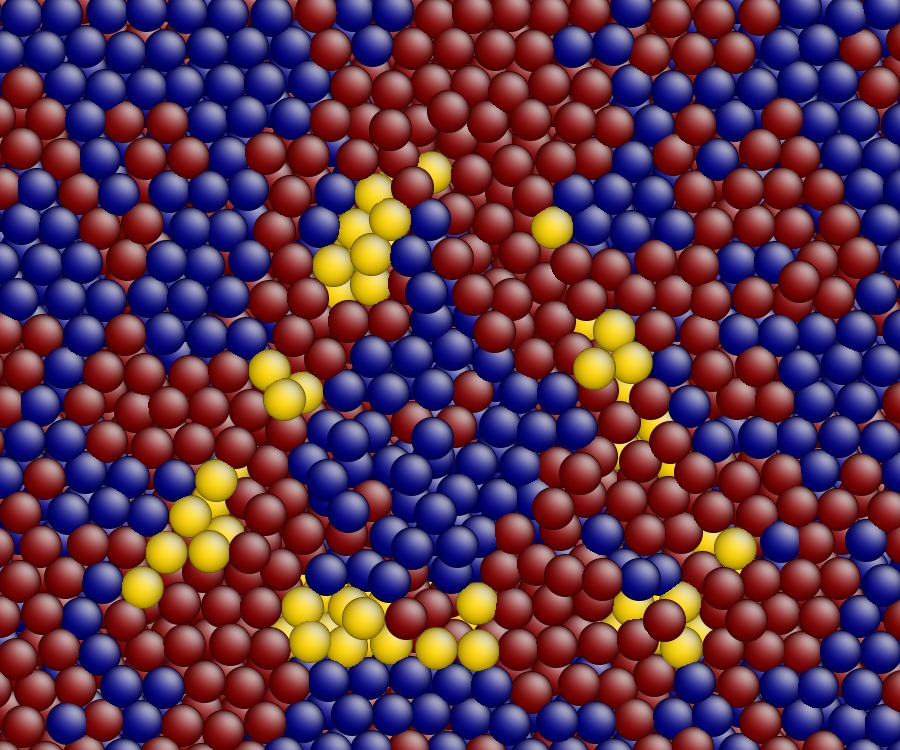} &
\includegraphics[width=0.47\columnwidth]{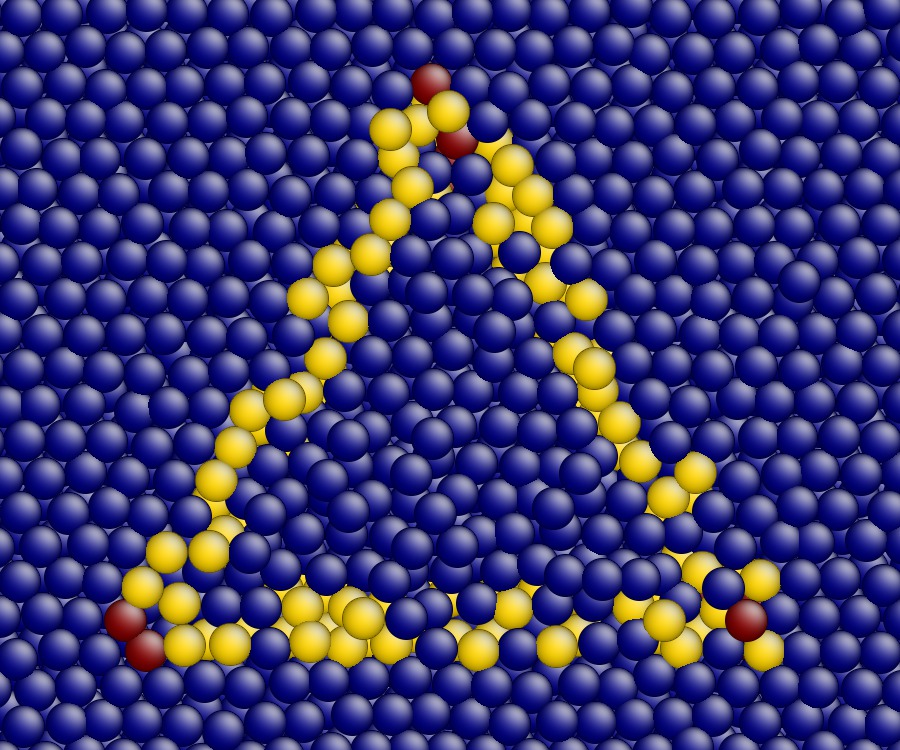} \\
(c) Adaptive CNA & (d) Voronoi Topology
\end{tabular}
\caption{Cross-section of a stacking-fault tetrahedron in copper at 85\% of its melting temperature, colored using several popular visualization approaches, and the proposed one.  In (b) and (c), dark blue, yellow and red indicate atoms in FCC, HCP, and other local environments, respectively.  In (d), dark blue, yellow, and red indicate atoms that are FCC-types, HCP- but not FCC-types, and all other types. \label{sft}}
\end{figure}

As a concrete example, we consider a stacking-fault tetrahedron \cite{kiritani1997story} in a high-temperature FCC aluminum single crystal.  A stacking-fault tetrahedron (SFT) is a three-dimensional defect consisting of four stacking faults that form the faces of a tetrahedron.  The interior of an SFT is  an FCC crystal; its edges are stair-rod dislocations \cite{hirth1982theory}.  Figure \ref{sft} illustrates a cross-section through the center of an SFT and parallel to one of its four faces; the intersection of the SFT boundary with the viewing plane is an equilateral triangle.  This perfect SFT was constructed in FCC copper and then thermalized at 0.85$T_{\text{m}}$.  Centrosymmetry, bond-angle, and adaptive common-neighbor analyses were all performed using the OVITO software package \cite{stukowski2010visualization}. 

Figure \ref{sft}(a) shows atoms colored using the centrosymmetry order parameter.  In this coloring, atoms belonging to faces of the SFT have higher centrosymmetry values than those in the FCC environment, as expected.  Note, however, that many atoms inside and outside the SFT also have high centrosymmetry values. Such atomic environments do not, however, indicate crystal defects, but rather result from thermal fluctuations which locally distort the atomic environment.  The inability of the centrosymmetry order parameter to distinguish structural defects from thermal perturbations requires users to quench a system before analyzing it.  

Figures \ref{sft}(b) and (c) show atoms colored using bond-angle analysis and adaptive common-neighbor analysis, respectively, also at 0.85$T_{\text{m}}$.  In these figures, many atoms belonging to the SFT faces are classified as having HCP local structure, as expected.  However, both methods identify many atoms away from the stacking faults as structural defects, despite the absence of other defects in the crystal. Moreover, application of bond-angle analysis incorrectly identifies many atoms in the bulk as having HCP local structure.  Although the general shape of the SFT can be discerned in both figures, the details are ambiguous and automated location of the SFT in an atomic ensemble is difficult or impossible at the simulation temperature.

These results are in contrast with the picture produced using Voronoi topology, illustrated in Figure \ref{sft}(d).  The approach taken here provides the clearest representation of the SFT.  In this case, every atom characterized as being in an HCP environment is on an SFT face, without exception, despite the high temperature and the strain fields of the constituent partial dislocations.  Moreover, all  atoms not at the surface of the SFT  are correctly identified as being FCC-type.  Finally, atoms lying at the corners of the triangular cross-section through the SFT triangle are identified as neither stacking faults (HCP-type) nor bulk type, but as having a distinct local structure; these are the dislocation cores.  The sole weakness of this visualization procedure results from indeterminate types which belong to both $\mathcal F_{\text{FCC}}$ and $\mathcal F_{\text{HCP}}$ and whose neighborhoods are identified as FCC rather than HCP.  This limitation can be addressed and is discussed in the supplementary material.  

This topological approach to structure visualization can also be applied to low-temperature systems such as those obtained through quenching (inherent structures); an example can be found in the supplementary material.

\section{Grain-Boundary \\Characterization and Analysis}
The proposed framework also enables the analysis of structurally-complex systems in an automated manner.  To illustrate this capability we analyze how a particular grain boundary transforms between a series of distinct structures as a result of absorbing point defects, as it may, for example, under irradiation conditions.
\begin{figure}[h]
\begin{tabular}[c]{ccc}
\fbox{\includegraphics[width=0.31\columnwidth,trim={0 3cm 0 2.75cm},clip]{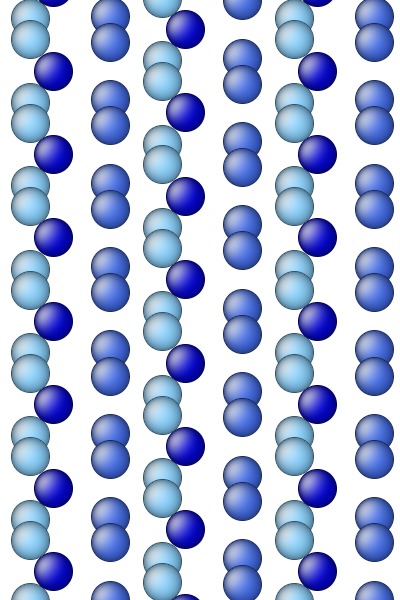}}  & 
\fbox{\includegraphics[width=0.31\columnwidth,trim={0 3cm 0 2.75cm},clip]{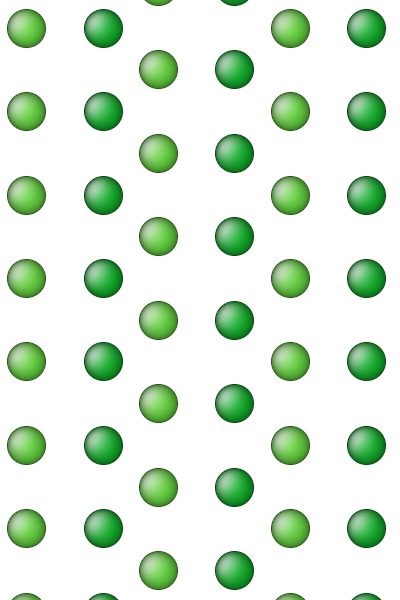}}  & 
\fbox{\includegraphics[width=0.31\columnwidth,trim={0 3cm 0 2.75cm},clip]{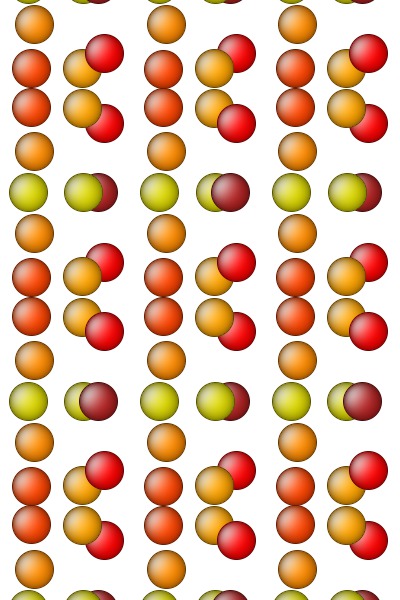}} \\  
(a) Phase I & (b) Phase II & (c) Phase III \vspace{1mm}\\
\end{tabular}
\begin{tabular}[c]{c}
\begin{overpic}[width=0.98\columnwidth]{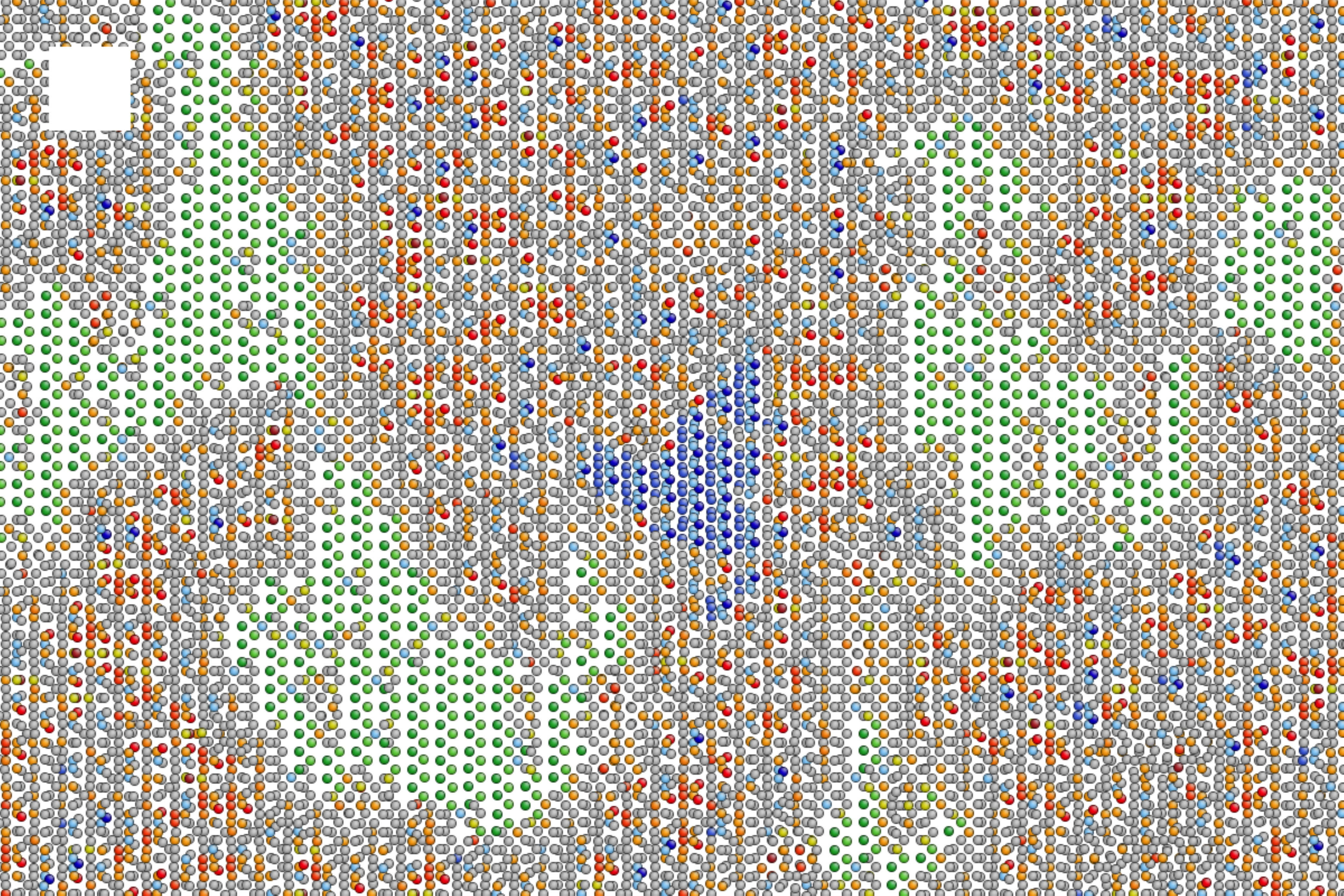}
 \put (4.35, 59.2) {(d)}
\end{overpic}
\end{tabular}
\caption{(a-c) Three structurally distinct stable phases of the $\Sigma 5$ [001] (310) symmetric tilt boundary in BCC tungsten, arbitrarily labeled Phase I, Phase II, and Phase III.  Atoms are colored according to topological type, each assigned a unique color; BCC-type atoms are not shown for clarity.  (d) An image from a large simulation of the same boundary, initially constructed uniformly of the Phase I structure, and into which self-interstitial atoms have been randomly placed at a constant rate at 1500 K.  Atoms are colored by topological type, as shown in (a-c); BCC-type atoms are not shown for clarity; all other atoms are shown in grey.\label{sigma5}}
\end{figure}

In particular, we consider a $\Sigma 5$ [001] (310) symmetric tilt boundary in a BCC tungsten bicrystal.  This grain boundary exhibits three structurally distinct, stable phases.  We begin by characterizing these phases using the Voronoi topologies of the constituent atoms; these phases are illustrated in Fig.~\ref{sigma5}(a-c).  Atoms are colored according to their topological type; BCC-type atoms are not shown for clarity.  Phase I consists of three distinct topological types, colored in different shades of blue, Phase II consists of two distinct topological types, colored in shades of green, and Phase III consists of six topological types, colored in shades of red, orange, and yellow.   

We initialize the simulation by constructing a large bicrystal containing a $\Sigma 5$ [001] (310) symmetric tilt boundary at 0 K and uniformly of the Phase I structure.  The sample is then heated to 1500 K, and equilibrated at this temperature for 4 ns.  Self-interstitial atoms are then inserted into random locations in the boundary plane at a rate of 1.62 atoms/nm$^2$/ns.  We analyze the resulting grain boundary structure throughout the MD simulation using Voronoi topology.  Figure \ref{sigma5}(d) shows a grain boundary with distinct domains of all three grain boundary phases, suggesting a phase transition driven by absorption of self-interstitial atoms.  

To study the transformation of the grain boundary structure, we track the fraction of each phase present during the evolution.  We begin by counting the number of atoms in the sample with topological types associated with each of the three phases.  We next calculate the number of non-BCC-type atoms per nm$^2$ in each of the three phases.  Finally, we normalize the $\lambda$-type counts for the three phases by dividing by the number of $\lambda$-types per unit area and the total grain boundary area.
 
\begin{figure}[h]
\includegraphics[width=1.\columnwidth]{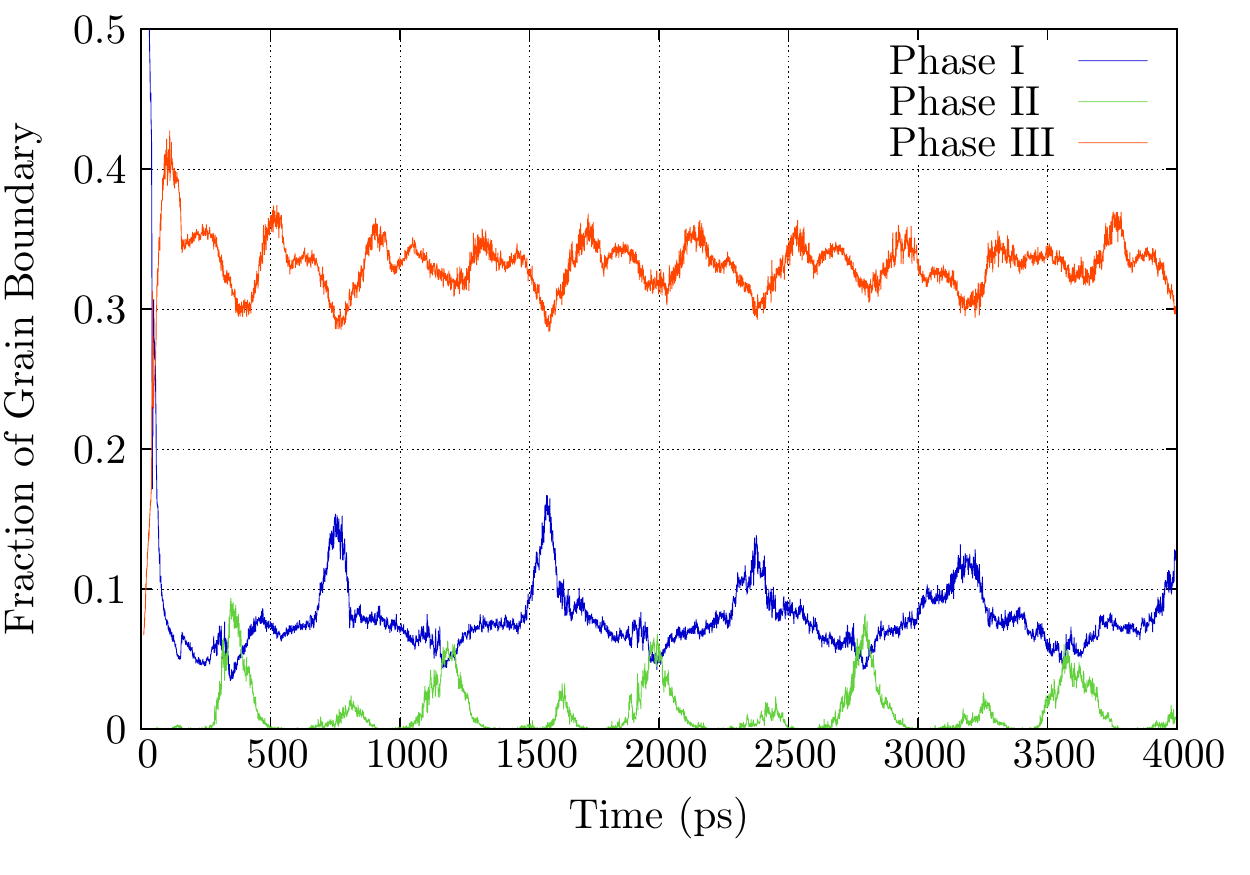} 
\caption{The fraction of the $\Sigma 5$ [001] (310) symmetric tilt boundary in BCC tungsten occupied by its structurally distinct, stable phases during the insertion of self-interstitial atoms at a fixed rate.\label{b123}}
\end{figure}

Figure \ref{b123} shows the fraction of the three phases over time, starting when the first self-interstitial atom is added to the grain boundary.  During the initial 100 ps, there is a sharp decrease in Phase I, accompanied by a rapid growth of Phase III.  After approximately 300 ps, the grain boundary structure settles into a pattern of increasing and decreasing Phase I, Phase II and Phase III fractions, all with the same period.  The minimum in the Phase III fraction corresponds to the maximum in the Phase I fraction and the maximum in the Phase II fraction corresponds to minima in the Phase I and III fractions.  The period is commensurate with the time required to add a full (310) plane of atoms to the sample.  At this temperature, the equilibrium grain boundary structure is dominated by Phase III, Phase I (the equilibrium structure at 0 K) never completely disappears, and Phase II is present only in a very small fraction of the grain boundary.

This example illustrates the power of the topological approach for automating structure analysis in systems with complex defect structures and for determining defect statistics.  We defer a more complete analysis of this example to a future study of grain boundary structure evolution during irradiation.

\section{Disordered Structures}

Finally, we consider how the proposed approach can be used to characterize disordered systems such as liquids and glasses.  In contrast to conventional order parameters -- which are typically useful for studying either ordered or disordered systems, but not both -- the approach taken here can be applied effectively to all kinds of systems.  As the topological type of each Voronoi cell provides a robust structural description of the local neighborhood of a particle, the distribution of topological types in a system can be interpreted as a statistical-topological description of the system as a whole.  This ability to characterize both ordered and disordered systems within the same framework is of particular importance for studying phase transitions between ordered and disordered phases, as well as between distinct disordered phases.  

Using MD, we simulate two disordered systems of copper atoms: a high-temperature liquid (HTL) equilibrated at roughly $1.85T_{\text{m}}$, and a glass-forming liquid (GFL) obtained by undercooling the initial liquid to roughly $0.75T_{\text{m}}$; each system contained 1,372,000 atoms.  The distributions of topological types in the two systems enable us to describe structural features of the systems in a robust and quantitative manner, and to observe structural differences between them.  

\begin{figure}[h]
\begin{tabular}[c]{c}
\includegraphics[width=1.0\columnwidth]{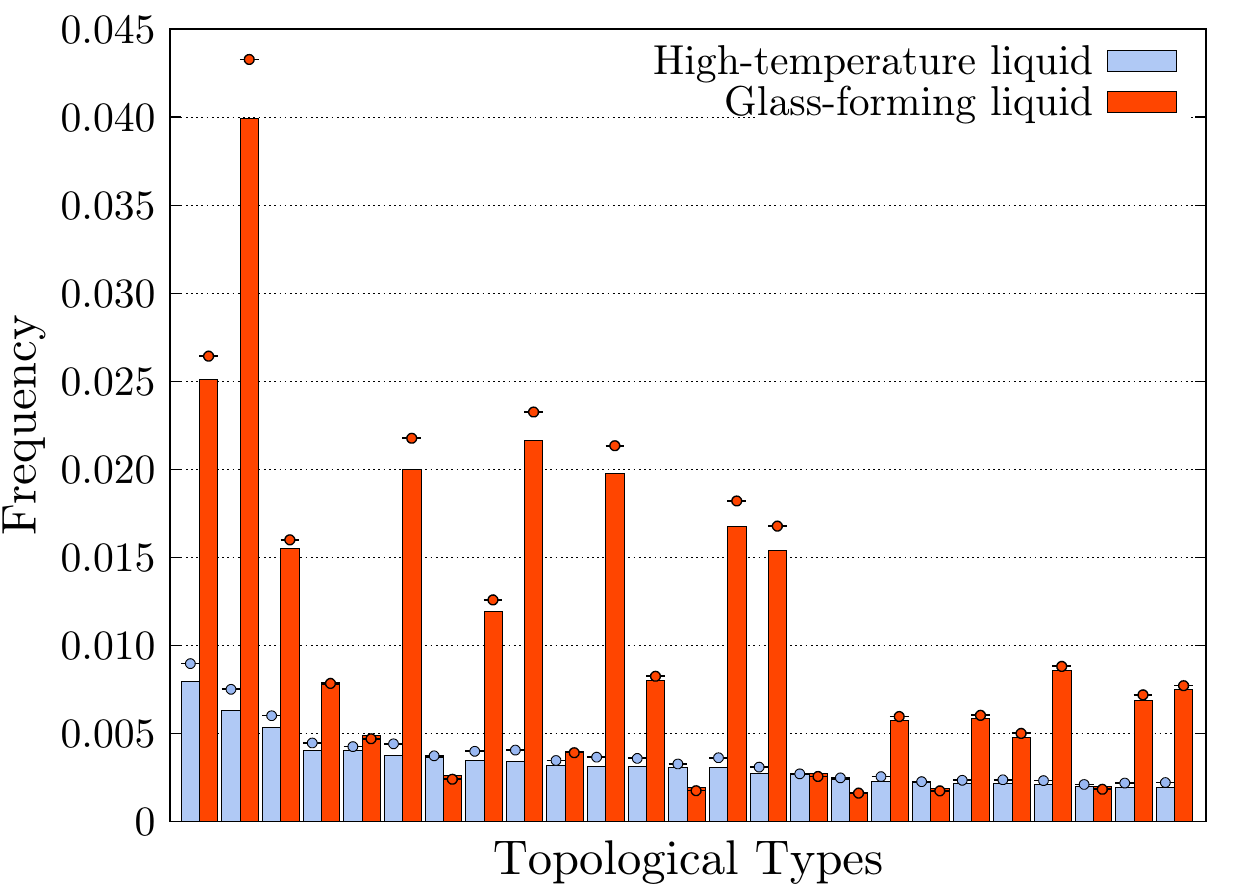} 
\end{tabular}
\caption{Frequencies of the twenty-five most common topological types in liquid copper at $1.85T_{\text{m}}$, and corresponding frequencies in the glass-forming liquid copper at $0.75T_{\text{m}}$; circles indicate frequencies in quenched samples.\label{glassetc}}
\end{figure}

Bars in Figure \ref{glassetc} show frequencies of the fifty most common topological types in the HTL, and corresponding frequencies in the GFL.  The most common types in the HTL do not occur nearly as frequently as the most common types in the GFL.  In particular, the sum of frequencies of the five most common topological types in the GFL is greater than the sum of frequencies of the fifty most common types in the HTL.  In this sense, the GFL is substantially more ordered than the HTL.  Note that in an ideal gas, in which there is no interaction between particles, the distribution of types is considerably less concentrated than in either of these systems \cite{2013lazar}.

Circles in Figure \ref{glassetc} indicate corresponding frequencies in systems obtained by quenching these two systems.  When quenched, frequencies of these common types change only modestly.  This again indicates that the current method is relatively insensitive to thermal vibrations.  

Other means of quantifying disorder in atomic systems have been widely developed, and have been used to distinguish distinct types of disordered systems.  Structural correlation functions \cite{truskett2000towards} and configurational entropy \cite{krekelberg2009generalized, goel2008tuning, goel2009available} are both non-local descriptors that have been used to study disordered systems.  Other recent work has focused on local structural measures \cite{bernal1959geometrical, sheng2006atomic}; our approach is in this vein.

\section{Conclusions and Discussion}
We have introduced a mathematical approach towards classifying and identifying local structure of a particle within a system of particles.  Applications highlight its utility in analyzing atomistic data sets, such as those produced by molecular dynamic simulations.  In particular, the theory of $\lambda$-types enables identification and visualization of defects in ordered systems at high temperatures.  This capability can play an important role for {\it in situ} study of high-temperature mechanisms currently inaccessible to current structure-identification methods.  The proposed framework also enables a new approach for characterizing and identifying defects.  This in turn allows for an automated approach for studying systems in which structural features evolve.  Finally, Voronoi topology enables the characterization of disordered systems in a statistical manner, through the distribution of topological types.  We have illustrated the potential of this approach in distinguishing a high-temperature liquid from a glass-forming liquid.  

Any description of structure within a fixed distance of a particle will be unable to capture all long-range structural features of a system.  Figure \ref{polycrystal} provides clear examples of this limitation, where an atom with HCP local environment might be part of a twin boundary, stacking fault, or other defect; further analysis is required to distinguish between these.  The analysis of local structure introduced here can be integrated into tools such as those developed in \cite{stukowski2012automated} to automate long-range structural analysis.

The authors have developed software to automate this analysis, and is available upon request.

\section*{Materials and Methods}

Deciding whether two Voronoi cells have the same topological type is equivalent to deciding whether two planar graphs are identical, as the edge-boundary of every Voronoi cell is a planar graph.  For each particle in a system we compute a ``code'' that records the graph structure of the edge-boundary network of its Voronoi cell.  To do this, we first determine the Voronoi cell using the {\footnotesize \tt Voro++} software package \cite{2009rycroft}, which computes a list of faces, each represented as an ordered sequence of vertices.  Next, we use a graph-tracing algorithm to compute a code for this planar graph.  More specifically, the following algorithm of Weinberg \cite{1966weinberg1} is followed: (a) An initial vertex is chosen and assigned the label 1. (b) An edge adjacent to that vertex is chosen and travel begins along that edge. (c) If an unlabeled vertex is reached, it is labeled with the next unused integer and we ``turn right'' and continue. (d) If a labeled vertex is reached after traveling along an untraversed edge, we return to the last vertex along the same edge but in the opposite direction. (e) If a labeled vertex is reached after traveling along an edge previously traversed in the opposite direction, we ``turn right'' and continue; if that right-turn edge has also been traversed in that direction, we turn along the next right-turn edge available; if all outgoing edges have been traversed, we stop.  At this point, each edge in the graph has been traversed once in each direction; the ordered list of the vertices visited is called a code.  

Codes are constructed for each choice of initial vertex and edge, and for each of two spatial orientations; all labels are cleared before producing each code.  For a Voronoi cell with e edges, 4e codes are generated, each an ordered list of 2e integer labels.  Each code completely describes the Voronoi cell topology, and so it is sufficient to only record one of them.  A code for a typical Voronoi cell requires less than 100 bytes of storage.  Additional details can be found in \cite{lazar2011evolution,1966weinberg1, 1966weinberg2}.
 
{\bf Run-time.} The use of Voronoi topology for structure identification is computationally efficient.  In preliminary tests, the Voronoi topology of one millions atoms could be calculated on a desktop computer in less than one minute.  By contrast, conventional approaches used in high-temperature structure analysis require that systems be quenched before visualization.  A complete quench necessary to obtain the inherent structure can require several hours of computation for a system of comparable size.

\begin{acknowledgments}
We gratefully acknowledge discussions with and assistance from Chris H. Rycroft and Zhaoxuan Wu.   Figures \ref{polycrystal}, \ref{sft}, and \ref{sigma5} were created with AtomEye \cite{li2003atomeye}.  EAL and DJS acknowledge support of the NSF Division of Materials Research through Award 1507013.
\end{acknowledgments}

\setcounter{figure}{0} 
\renewcommand{\thefigure}{S\arabic{figure}}

\onecolumngrid 
\vspace{15mm}
\begin{center}
\textbf{\large SUPPLEMENTARY MATERIAL}
\end{center}

\vspace{8mm}

\twocolumngrid

\section*{Enumerating Primary \\and Secondary Types} 
In the paper we considered families of topological types $\mathcal F_{\lambda}$ associated with particular structures.  In this section we provide some additional detail regarding the determination of these families, and report the numbers of types in several of them, as well as in the overlap between multiple families.

The Voronoi cell of BCC is topologically stable in the sense that infinitesimal  perturbations of atomic coordinates will not change its topology.  Therefore, $\mathcal F_{\text{BCC}}$ consists of a single primary type and no secondary types.  

In contrast, Voronoi cells of FCC and HCP are unstable, and infinitesimal perturbations of atomic coordinates will change their topologies.  The instability of these Voronoi cells can be detected in vertices that are incident with four edges; there are six such unstable vertices in FCC and HCP.  As each unstable vertex can either remain unstable, resolve in one of 3 primary directions, or resolve in one of 4 secondary directions (see Fig.~3), we must check $8^6 = $ 262,144 configurations that can result from all infinitesimal perturbations.  We compute the topology of each configuration using the algorithm described in the Materials and Methods section of the paper.  Multiple configurations can result in the same Voronoi cell topology due to symmetries of the unperturbed configuration.

\begin{figure}[h]
\begin{tabular}[c]{ccc}
\includegraphics[width=0.27\columnwidth]{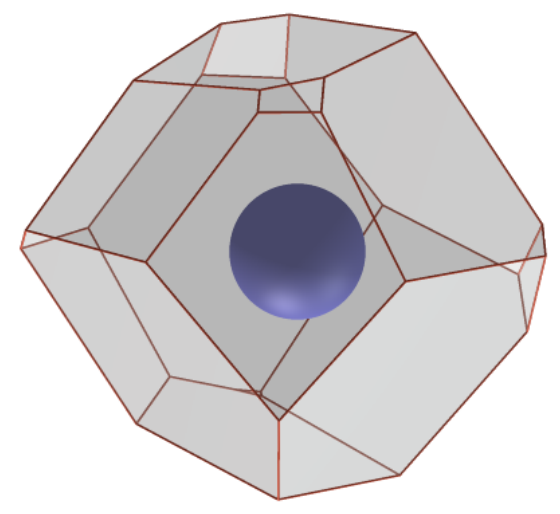} &
\includegraphics[width=0.27\columnwidth]{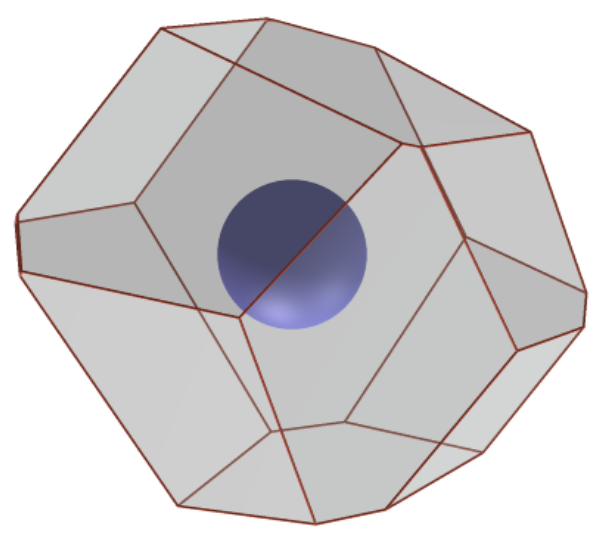} &
\includegraphics[width=0.27\columnwidth]{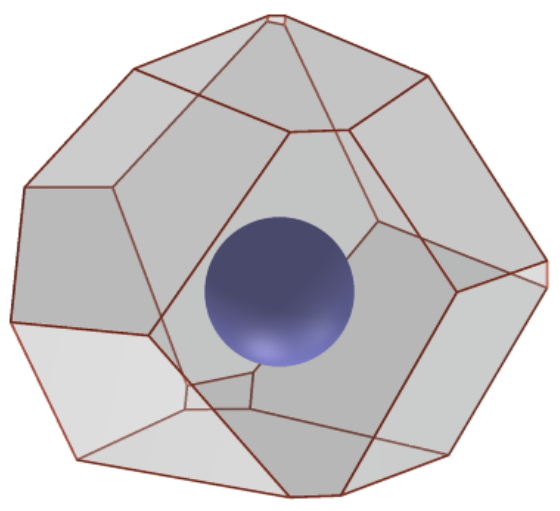} 
\end{tabular}
\caption{Examples of Voronoi cells in a perturbed FCC crystal.\label{perturbations}}
\end{figure}

For FCC we find 44 primary types and 6250 secondary types; of the secondary types only 2771 have no unstable vertices.  Figure \ref{perturbations} illustrates several Voronoi cells observed in a finite-temperature FCC crystal; their topologies are given by this enumeration technique.  For HCP we find 66 primary types and 21,545 secondary types; of the secondary types only 9490 have no unstable vertices.

While the determination of topological types associated with a particular structure may require substantial computation, this needs only be done once per structure.  Lookup tables are then created and subsequently referenced when analyzing atomistic data sets.  Data for several common crystal structures are available from the authors upon request.

In the paper we noted that certain types can belong to several families (indeterminate types).  More specifically, we note that the unique type belonging to $\mathcal F_{\text{BCC}}$ also belongs to $\mathcal F_{\text{FCC}}$ and $\mathcal F_{\text{HCP}}$.  Furthermore, $\mathcal F_{\text{FCC}}$ and $\mathcal F_{\text{HCP}}$ share 23 primary types in common and 1352 secondary types in common.

\section*{Resolving Indeterminate Types} 
In the paper we noted practical challenges created by the overlap of multiple families of types.  In particular, when attempting to visualize defects in an FCC environment, some atoms were mistakenly identified as belonging to the FCC bulk instead of to the HCP-like defect.  Although a thorough analysis of this problem is beyond the scope of this paper, we briefly consider one approach towards resolving it.  We leave a complete discussion of this topic for a future paper.

Indeterminate types are Voronoi topologies that can be obtained through infinitesimal perturbations of multiple distinct structures.  Figure 2 in the paper shows a number of regions in configuration space incident with both $\mathbf x_{\text{FCC}}$ and $\mathbf x_{\text{HCP}}$.  One way of deciding whether a point in this region should be classified as FCC or HCP involves computing distances in this configuration space: points close to $\mathbf x_{\text{FCC}}$ should be classified as FCC, while those close to $\mathbf x_{\text{HCP}}$ should be classified as HCP.  Points closer to $\mathbf x_{\text{BCC}}$ than to either $\mathbf x_{\text{FCC}}$ or $\mathbf x_{\text{HCP}}$ should be classified as BCC.

While theoretically appealing, there is no practical manner in which to calculate these distances because of the complicated topology of this configuration space.  More specifically, the standard metric on $\mathbb{R}^{3n}$ cannot be used due to the actions of the rotation, renormalization, and permutation groups acting on it.

A more practical approach involves perturbing configurations of particles as follows.  If the Voronoi type of a particle is indeterminate, we randomly perturb the particle and its nearby neighbors; this corresponds to a small perturbation of $\mathbf x$ in configuration space.  The Voronoi cell of the perturbed configuration is calculated; this procedure is repeated several times.   If all perturbations result in indeterminate or HCP-types (the latter occurring at least once), then the particle is classified as being HCP-type.
\begin{figure}
\begin{tabular}[c]{cc}
\includegraphics[width=0.47\columnwidth]{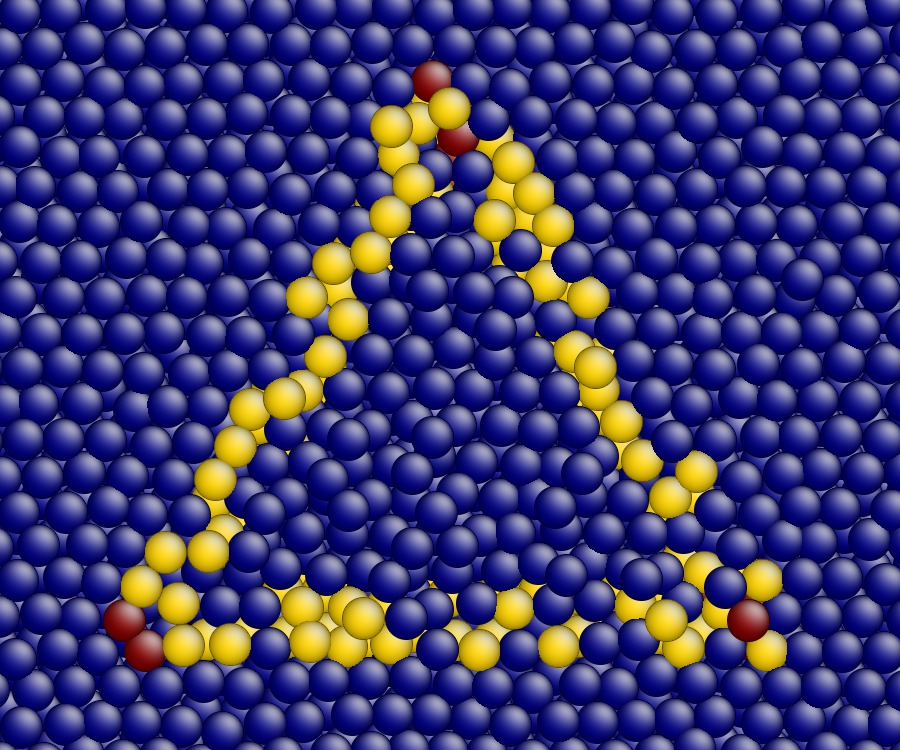} &
\includegraphics[width=0.47\columnwidth]{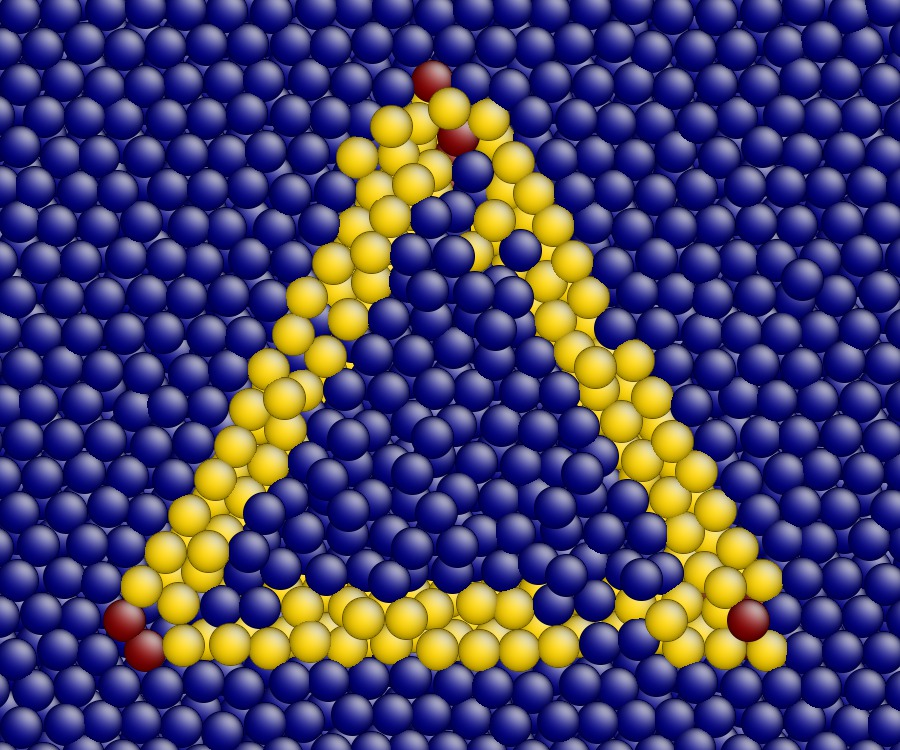} \\
(a) & (b) 
\end{tabular}
\caption{A high-temperature single crystal containing an SFT, visualized using (a) the method described in the paper and (b) the modified method proposed here.\label{einstein}}
\end{figure}

We consider a stacking-fault tetrahedron (SFT) in a single copper crystal that was heated to 85\% of its melting temperature; this system was considered in ``Comparison with Other Methods'' section of the paper.  Figure \ref{einstein} shows the SFT colored by the approach considered in the paper and by the modification considered here.  This modified approach provides a more robust visualization of the SFT than the approach suggested in the paper.  We defer a complete discussion of indeterminate types and methods of resolving them to a future paper.

\section*{Quantitative Comparison \\at High Temperatures} 
In the paper we have shown that Voronoi topology enables robust visualization of structure in high-temperature systems that cannot be obtained using conventional methods.  Here we provide a direct and quantitative comparison beyond visual inspection.  We begin with a system containing 1,372,000 copper atoms organized in a perfect crystal, and heat the system to just below its melting point as described in the ``Finite-Temperature Crystals'' section of the paper.  For each order-parameter considered in the paper, we calculate the frequency of atoms in this single crystal that are characterized as non-FCC-type as a function of temperature.  Close inspection of the system shows that there are no defects.

\begin{figure}[h]
\begin{tabular}[c]{c}
\includegraphics[width=1.0\columnwidth]{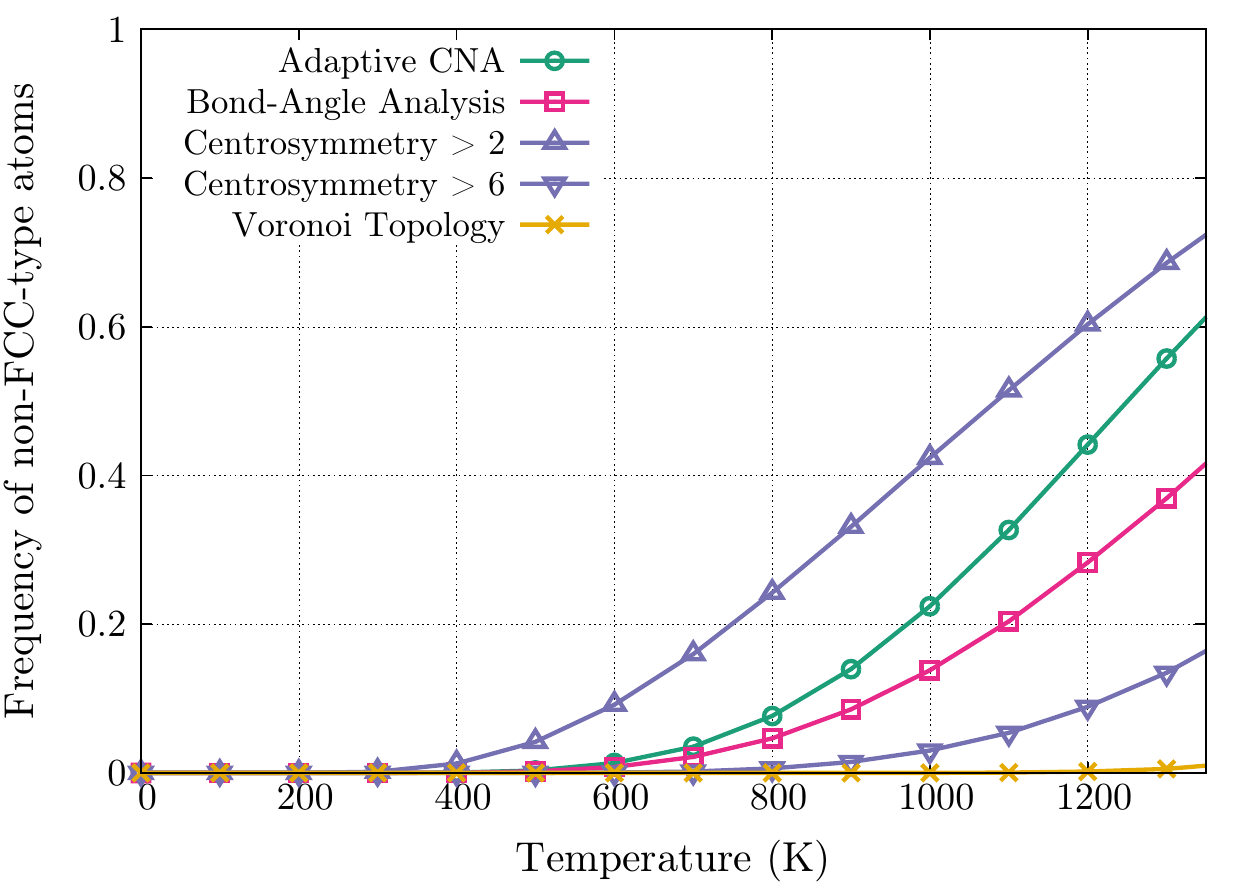}
\end{tabular}
\caption{Frequency of atoms characterized as non-FCC-type in a copper crystal.\label{defects}}
\end{figure}

For each visualization method we considering the number of atoms in this system characterized as non-FCC-type.  We use OVITO to compute bond-angle analysis, adaptive common-neighbor analysis, and centrosymmetry values for each atom and at each temperature.  Figure \ref{defects} shows the frequency of non-FCC-type atoms, as measured by each of the order-parameters, as a function of temperature.  

In contrast to the bond-angle and adaptive common-neighbor classifications provided by OVITO, centrosymmetry (CS) is reported as a scalar requiring choice of a cutoff; atoms with CS values above this cutoff are not considered FCC.  To determine an appropriate cutoff we considered CS values of atoms adjacent to either a vacancy or an interstitial in an otherwise perfect FCC crystal.  Atoms adjacent to a vacancy have a CS value of 6.20; interstitial atoms have a CS value of 13.88.  In order that our choice of cutoff allows for the detection of vacancies, we choose a cutoff of 6.  We note that cutoff values as low as 0.5 and 1 can be regularly found in the literature, and so our choice of cutoff is extremely conservative.  

At $T = 1300$ K, bond-angle analysis, adaptive common-neighbor analysis, and centrosymmetry incorrectly identify over 36.9\%, 55.7\% and 13.5\%, respectively, of the atoms as not belonging to the FCC bulk crystal.  By contrast, Voronoi topology mistakenly identifies only $0.53\%$ as such atoms as not belonging to the FCC bulk crystal.

\renewcommand{\floatpagefraction}{.8}%

\section*{Quenched Systems} 
\begin{figure}
\begin{tabular}[c]{cc}
\includegraphics[width=0.47\columnwidth]{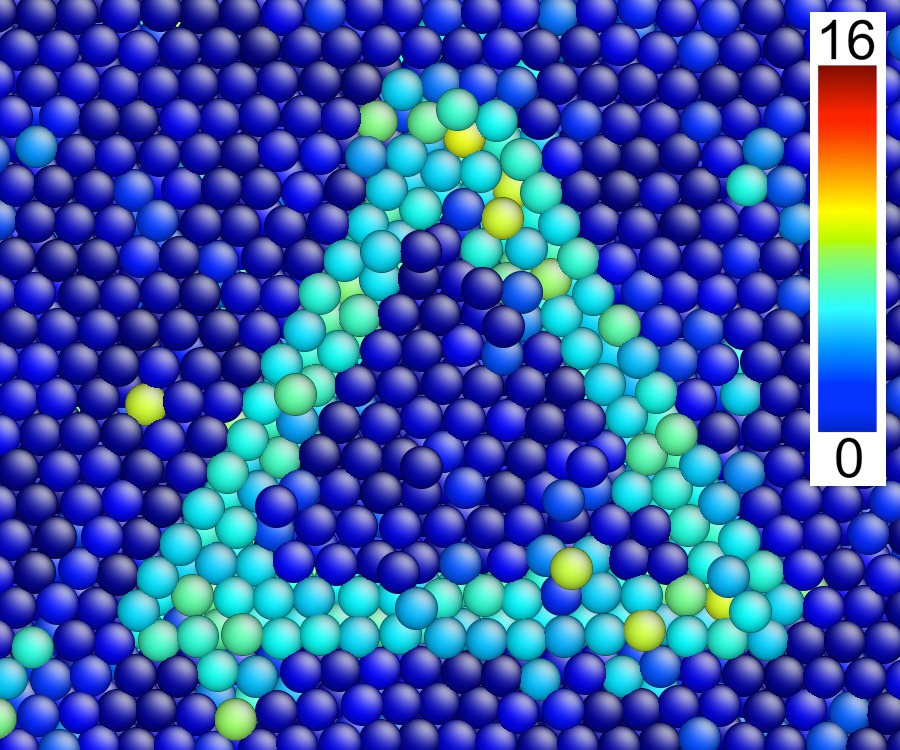} \quad &
\includegraphics[width=0.47\columnwidth]{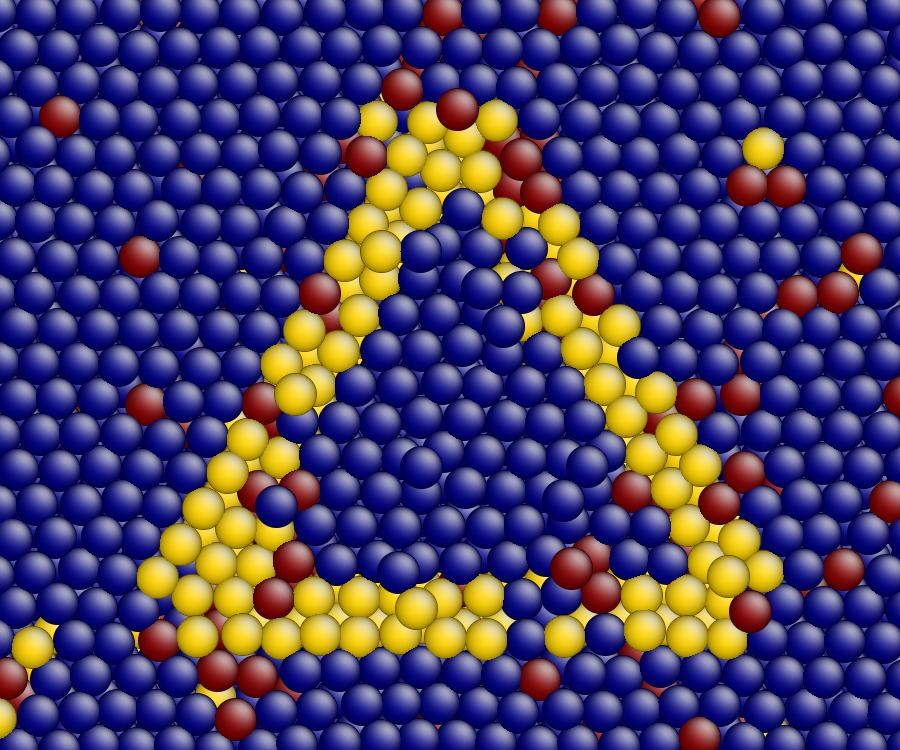} \\
(a) Centrosymmetry  & (b) Bond-angle analysis \vspace{2mm} \\
\includegraphics[width=0.47\columnwidth]{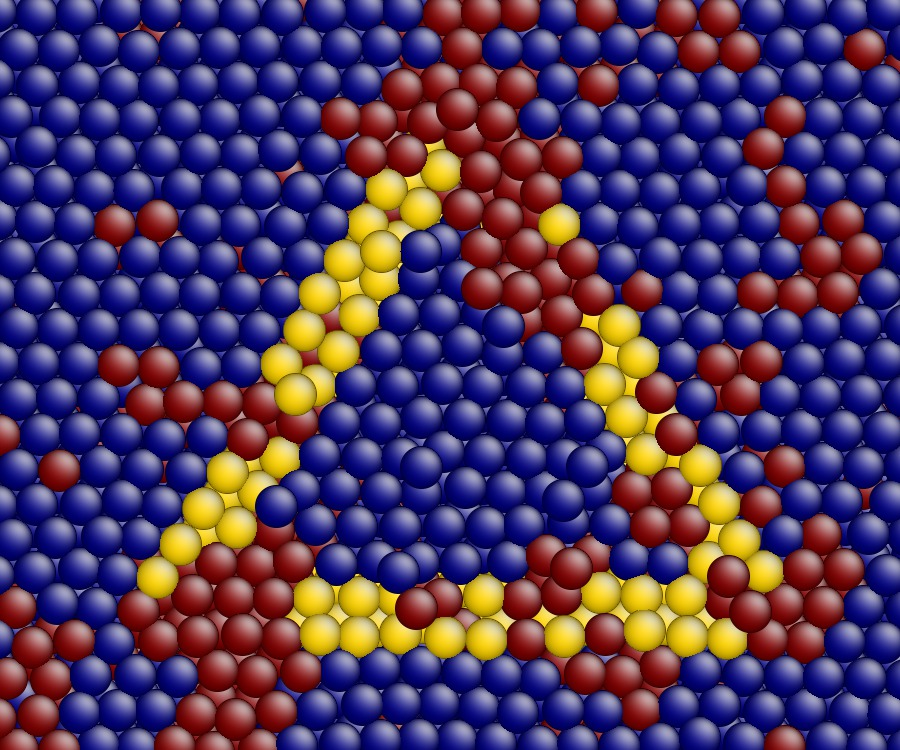} &
\includegraphics[width=0.47\columnwidth]{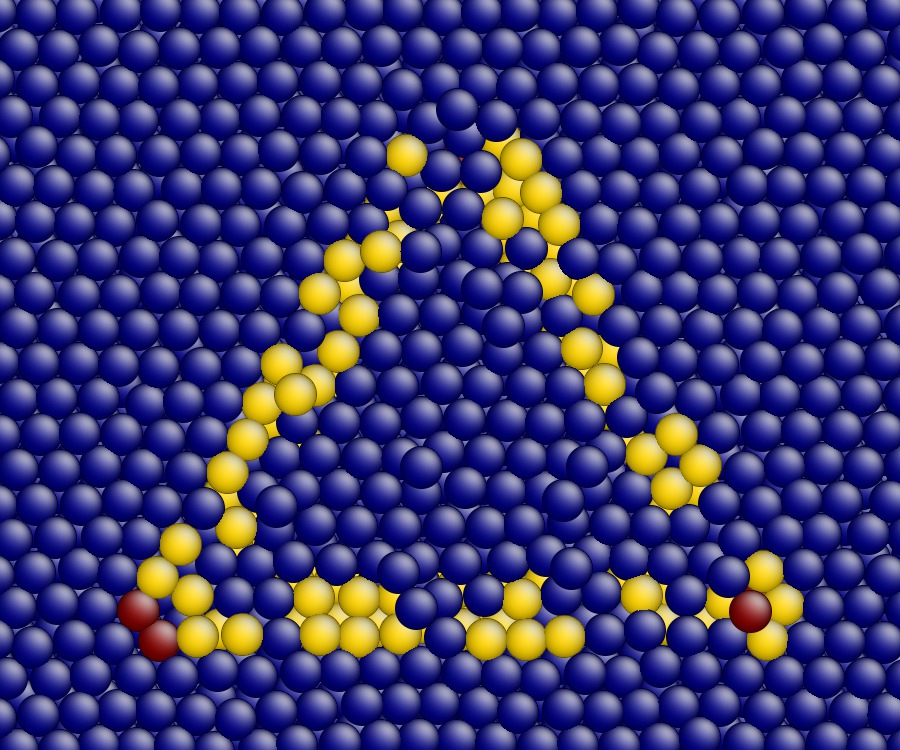} \\
(c) Adaptive CNA & (d) Voronoi Topology
\end{tabular}
\caption{Cross-section of a stacking-fault tetrahedron in copper, after quenching from 85\% of its melting temperature, colored using several popular visualization approaches, and Voronoi topology.  In (b) and (c), dark blue, yellow and red indicate atoms in FCC, HCP, and other local environments, respectively.  In (d), dark blue, yellow, and red indicate atoms that are FCC-types, HCP- but not FCC-types, and all other types, respectively.\label{sftq}}
\end{figure}

The topological approach is not limited to studying finite-temperature systems, and can also be applied to structures obtained through quenching high-temperature samples.  Figure \ref{sftq} shows the same stacking-fault tetrahedron (SFT) considered in the paper after quenching to 0 K, colored using the same order-parameters considered in the paper.

Although all methods successfully detect the presence of the SFT in the quenched system, centrosymmetry, bond-angle analysis, and adaptive common-neighbor analysis also identify atoms outside the SFT as being non-FCC-type.  These non-FCC-type identifications are the result of small local strains which are not structural defects.  Visualization of the sample through Voronoi topology does not result in any such misidentifications, providing the most robust picture of structural defects.
\vspace{3mm}

\FloatBarrier

\end{document}